\documentclass[a4paper,pre,twocolumn,superscriptaddress,nofootinbib]{revtex4-2}
\usepackage[english]{babel}
\usepackage[utf8]{inputenc}
\usepackage{graphicx,epsfig,xcolor}
\usepackage{latexsym}
\usepackage{amsfonts}
\usepackage{amsmath}
\usepackage{textcomp}
\usepackage{hyperref}
\usepackage{placeins}
\usepackage{stmaryrd}
\usepackage{hhline}
\usepackage{slashbox}

\usepackage{units}
\usepackage{amssymb}
\usepackage{bbm}
\usepackage{verbatim}
\usepackage{anysize}
\usepackage{array}
\usepackage{float}

\def\<{\langle}
\def\>{\rangle}
\def\{{\lbrace}
\def\}{\rbrace}
\def\({\left(}
\def\){\right)}
\def\beq{\begin{equation}}
\def\eeq{\end{equation}}

\def\Anu{A^\nu}

\def\wmax{w_{\text{max}}}

\def\lopt{\ell_{\text{opt}}}
\def\loptmin{\ell_{\text{opt}}^{\text{min}}}
\def\loptmax{\ell_{\text{opt}}^{\text{max}}}
\def\loptstar{\ell_{\text{opt}}^{\star}}

\def\loptmean{\overline{\ell}_{\text{opt}}}

\def\loptmeanc{\overline{\ell}_{\text{opt}\,\times}}
\def\loptmeancA{\overline{\ell}_{\text{opt}\,\times}^{(a)}}
\def\loptmeancB{\overline{\ell}_{\text{opt}\,\times}^{(b)}}

\def\rc{r_\times}

\def\Lsat{L_\text{sat}}

\def\dopt{d_{\text{opt}}}
\def\gopt{g_{\text{opt}}}

\def\wmax{w_{\text{max}}}

\begin{document}

\title{A unified scaling for the optimal path length in disordered lattices}

\author{Daniel Villarrubia-Moreno}
\affiliation{Dpto. Matem\' aticas \& Grupo Interdisciplinar de Sistemas Complejos (GISC), Universidad Carlos III de Madrid, Legan\'es (Spain)}

\author{Pedro C\'ordoba-Torres}
\affiliation{Dpto. F\'{\i}sica Matem\'atica y de Fluidos, Universidad Nacional de Educaci\'on a Distancia (UNED), Madrid (Spain)}

\date{January 18, 2024}

\begin{abstract}
In recent decades, much attention has been focused on the topic of optimal paths in weighted networks due to its broad scientific interest and technological applications. In this work we revisit the problem of the optimal path between two points and focus on the role of the geometry (size and shape) of the embedding lattice, which has received very little attention. This role becomes crucial, for example, in the strong disorder limit, where the mean length of the optimal path for a fixed end-to-end distance diverges as the lattice size increases. We propose a unified scaling ansatz for the mean length of the optimal path in $D-$dimensional disordered lattices. The ansatz is supported by a comprehensive numerical study of the problem on $2D$ lattices, yet we also present results in $D=3$. We show that it unifies well-known results in the strong and weak disorder regimes, including the crossover behavior, but it also reveals novel scaling scenarios not yet addressed. Moreover, it provides relevant insights into the origin of the universal exponents that characterize the scaling of the optimal path in the strong disorder limit.
\end{abstract}

\maketitle


\section{Introduction}
\label{sec:intro}

The geometry and dynamics of optimal paths in weighted networks have been of major interest in the past decades \cite{1,2}. This is due to two principal reasons. First, because they are closely related to important optimization problems such as: \emph{minimum spanning trees} \cite{3,4,5,6,7}, \emph{directed polymers in random media} (DPRM) \cite{8}, \emph{optimal polymers} (fixed-length \emph{self-avoiding walks}) \cite{9,10}, \emph{spin glasses} \cite{11,12,13}, \emph{first-passage percolation} \cite{14,15,15a}, or the \emph{traveling salesman problem} \cite{16}. Secondly, they play a fundamental role in relevant scientific and technical applications which include: magnetotransport in disordered thin films \cite{17,17b}, fluid flow through porous media \cite{18,19,20,21,22}, current flow in \emph{random resistor networks} \cite{23,24}, fracture or crack processes in random media \cite{25}, transport and routing in communication networks \cite{4,5,6,7}, traffic engineering \cite{6,7,26,27}, percolation models \cite{28}, and epidemic spreading \cite{29}. In general, the resources of real-world complex networks are most efficiently used when transport follows optimal paths \cite{30}.

The formulation of the problem is simple. A network is disordered by assigning a random \emph{weight} $w_i$ to each bond $i$ connecting two nodes of the network. The link weights are positive, independent and identically distributed random variables with probability distribution $F(w)$ and  density function $f(w)$. The independence of the weights is a good approximation for large systems, but in some real-world networks they might be correlated with the network topology \cite{31}.

The optimal path between two nodes is the path that minimizes the sum of the weights along the path, and its length $\lopt$ is given by the number of bonds along the path. Optimal paths have been studied both in regular lattices \cite{5,7,11,25,31,32,33,34,35,36} and in random networks, principally Erdös-Rényi and \emph{scale-free networks} \cite{4,5,6,7,24,37,38,39,40,41}.

In this work we are concerned with the lattice problem, so in this introduction we will focus on the main results related to disordered lattices, yet most of them apply in general. Also, unless we say the contrary, we will use $r$ to refer to the Euclidean distance (in lattice units) spanned by the optimal path. Thus, for the optimal path between two points, it is the end-to-end distance between them, while for the problem of the optimal path connecting two opposite edges of a lattice of linear size $L$, it is given by $L$.

On weakly disordered lattices, almost all links contribute to the total weight of the optimal path. Optimal paths in \emph{weak disorder} (WD) belong to same \emph{universality class} as DPRM \cite{33,34,42}, which can be mapped to the celebrated Kardar-Parisi-Zhang (KPZ) universality class \cite{8,14,43}. At large length scales the optimal paths are self-affine and $\lopt$ scales linearly with $r$, whatever the dimension $D$ of the embedding space \cite{32,33,34}.

\emph{Strong disorder} (SD) is obtained from extremely broad distributions. Then we can assume that the total weight of a path is completely dominated by the largest bond-weight along it, $\wmax$, so that the optimal path is the path with the minimum value of the maximum link-weight \cite{11,22,34,35,36,40}. We should stress that this assumption is rigorously valid only in the so-called \emph{strong disorder} (or ultrametric) \emph{limit} \cite{11,40}.

It is agreed that the scaling properties of optimal paths in the SD limit are also universal \cite{11,35,36} and that they belong to the same universality class as the paths connecting nodes in minimum spanning trees \cite{3,4,5,6,7} and in \emph{invasion percolation} clusters \cite{20,21,22}, although the latter case has been questioned \cite{44}.

Optimal paths in SD are self-similar objects. and their mean length $\loptmean$ scales with $r$ as $\loptmean \sim r^{\dopt}$ \cite{11,32,33,34,35,36}. Fractal dimension $\dopt$ is universal, i.e., it depends only on the dimension $d$ of the system \cite{36}, and seems to be a ubiquitous property of many natural systems \cite{45}. There are no exact values for $\dopt$, yet we have accurate approximates obtained from numerical simulations \cite{20,34,36}. They roughly agree in $\dopt= 1.22 \pm 0.02$ for $D=2$, and $\dopt=1.42 \pm 0.03$ for $D=3$.

The behavior of the optimal path in the SD limit is closely related to critical \emph{percolation} \cite{11,22,40}, so the application of this theory has led to many relevant results \cite{32,33,34,35,36}. The optimal path in the SD limit belongs to the backbone of the bond-percolation cluster obtained at probability $p=F(\wmax)$. For $r=\infty$ we expect $\wmax=F^{-1}(p_c)$, where $p_c$ is the critical percolation threshold of the bond-percolation problem in the same lattice \cite{46}. For a finite $r$, $\wmax$ is distributed around $F^{-1}(p_c)$ \cite{33,34} with a standard deviation that scales as $\sim r^{-1/\nu}$ \cite{33}, where $\nu$ is the percolation correlation length exponent \cite{46}.

Another relevant result concerns the distribution of the optimal path length,  $P(\lopt)$, in the SD limit \cite{35,36}. For optimal paths between two sites separated a distance $r$ on a $D$-dimensional square lattice of linear size $L\gg r$, $P(\lopt)$ seems to follow a power-law decay between a lower and an upper cutoff length, with an exponent $-\gopt=-1.55\pm 0.05$ in $d=2$ and $-\gopt=-1.37\pm 0.05$ in $d=3$ \cite{35}. The upper cutoff is a finite size effect and its value seems to increase with $L$. This result leads to an important  conjecture that is verified here: $\loptmean$ and all higher moments diverge as $L\rightarrow \infty$.

For intermediate disorder, optimal paths behave as in the SD limit over small distances, and as in WD at large length scales. Thus, a crossover from self-similar to self-affine behavior is obtained in both regular \cite{32,33,34} and random graphs \cite{37,38,39,40,41} which also affects the properties of the global transport \cite{4,5,6}. Similar disorder-induced transitions have been reported in related models such as Ising systems \cite{11,12,13}, fixed-length self-avoiding walks \cite{10}, and in DPRM \cite{47}.

A general approach for the crossover length scale, denoted here by $\rc$, was proposed in Ref. \cite{32}. It was shown that $\rc$ scales with disorder as $\rc\sim  A^\nu$, where $A$ is a measure of the disorder strength that depends on the weight distribution and on $p_c$. The model generalizes to arbitrary distributions the expression obtained for exponential disorder \cite{17,17b,23,33}. The critical percolation exponent $\nu$ appears again as a consequence of the close relationship between the SD limit and critical percolation. The same scaling is obtained for the crossover network-size in random graphs \cite{32,39,40,41}. Remarkably, the scaled parameter $r/\rc$ fully determines the properties of the optimal path length in both strong and weak disorder \cite{20,32,33,35,36}.

Accordingly, the length of the optimal path at the crossover point, denoted by $\loptmeanc$,  scales with disorder as $\loptmeanc \sim \rc ^{\dopt} \sim  A^{\nu \dopt}$ \cite{33}. Although this is the general belief, this result disagrees with some results reported in Ref. \cite{34}. The authors obtained $\loptmeanc \sim A^\kappa$ with exponent $\kappa\simeq1.60$ in both $d=2$ and $d=3$, so it seems to be independent of the dimensionality of the system. If we compare both exponents we get that $\nu \dopt \approx \kappa$ for $D=2$ ($1.63\approx 1.60$), but the equality does not hold in $D=3$ ($1.25\neq 1.60$). To the best of our knowledge, this disagreement has not been addressed.

While much effort has been devoted to understand the influence of disorder on the optimal path problem \cite{48}, the influence of the lattice geometry on the optimal path between two nodes of the lattice has gone practically unnoticed, with a few notable exceptions \cite{35}. The distance spanned by the optimal path, $r$, and the linear size of the lattice, $L$, have been usually considered to be equal, $r=L$ \cite{11,32,33}, or linearly related, $L\propto r$ \cite{36}. However, the issue turns out to be very relevant in the SD limit, where $\lopt$ seems to diverge as $L\rightarrow \infty$ for a fixed end-to-end distance $r$ \cite{35}.

The aim of this paper is to unify the results reported in the literature into a more general description that also accounts for the effects of the lattice geometry. For that purpose, we present a unified scaling ansatz for $\loptmean$ which is derived from a thorough numerical study of the problem and from some theoretical arguments. We study the different scaling scenarios included in our model, some of which are novel. From a fundamental point of view, it provides valuable insights into the universality accounting for the behavior in the SD limit.

This article is organized as follows. We briefly discuss in Sec. \ref{sec:model} the model and our basic assumptions. Next, in Sec. \ref{sec:scaling_ansatz} we present our unified scaling ansatz. The different scaling regimes derived from it are presented in Sec. \ref{sec:scaling-regimes} and discussed in Sec. \ref{sec:discussion}. We will pay special attention in comparing our results to those reported in the literature. Numerical results supporting our approach and illustrating the scaling regimes and their transitions will be shown in Sec. \ref{sec:num_results}. Finally, Sec. \ref{sec:conclusions} is devoted to a
summary of our conclusions and our ideas regarding future work.


\section{Model and definitions}
\label{sec:model}

We consider a $D$-dimensional simple hypercubic lattice of size $L_1\times L_2\times \cdots\times L_D$, and assign to each link $i$ of the lattice a random weight $w_i$. The link weights are assumed to be independent, identically distributed random variables, drawn from the probability distribution $F(w)$ with density function $f(w)$, and they are strictly positive, $F(0) = 0$. End points A and B are centered on the lattice along the direction $L_1$, and separated by a Euclidean end-to-end distance $r$. Notice that this implies $L_1\geq r$. For example, for square $L\times L$ lattices, points A and B are located at sites $(-r/2,0)$ and $(r/2,0)$, while the corners of the lattice are sites $(\pm L/2, \pm L/2)$. Then we apply Dijkstra’s algorithm \cite{49} to find the optimal path between the two nodes. Its length $\lopt$ is given by the number of links along the path (we set the distance between neighbouring points on the lattice equal to unity).

Two different weight distributions have been usually employed in the related literature for generating a broad disorder.  The first one is the inverse distribution $f(w)=1/(aw)$ with $w\in[1,e^a]$ \cite{32,33,34,35,36,37,38,39,40,41}. The positive parameter $a$ controls the broadness of the disorder so it can be considered as a direct measure of the disorder strength. The limit $a\rightarrow \infty$ is the SD limit, where the largest $w_i$ along the path dominates the sum. The limit $a\rightarrow 0$ is the WD limit (\emph{homogeneous} case), where all links have the same weight, so they contribute equally to the sum. In the WD limit, the optimal path is given by the shortest path.


Another distribution that has also been widely used to generate disorder is the polynomial distribution $F(w)=w^\alpha$ with $\alpha >0$ and $w\in[0,1]$ \cite{4,5,6,7}. It is the simplest distribution with a distinct different behavior for small values of $w$ than regular distributions \cite{4,5,6,7}. The exponent $\alpha$ is called the \emph{extreme value index} and controls the disorder strength. The SD and WD limits are obtained when $\alpha\rightarrow 0$ and $\alpha\rightarrow \infty$, respectively.

Since universal behavior is expected to be independent of the type of distribution \cite{3,32,14,15}, in this work we have followed a different approach and considered an unbounded weight distribution, concretely the Weibull distribution, with probability density function
  \beq
    f(w)={k\over\lambda} \({w\over\lambda}\)^{k-1}\exp\(-(w/\lambda)^k\),
  \label{eq:wei_f_F}
  \eeq
where $\lambda>0$ and $k>0$ are the scale and shape parameters, respectively, and which is defined only for positive $w$.

As explained in the introduction, the SD-WD crossover end-to-end distance has the form \cite{17,17b,23,32,33}
\beq
\rc \sim \Anu.
\label{eq:rc}
\eeq
Here we use the definition of the \emph{disorder strength parameter} $A$ given in Ref. \cite{32}, $A= p_c/\left(w_c f(w_c)\right)$, where $w_c=F^{-1}(p_c)$. For the Weibull distribution we obtain
\beq
\label{eq:A}
A=\frac{p_c}{(p_c-1)\ln{(1-p_c)}k}.
\eeq

We note that, for a given lattice (hence a given $p_c$), disorder parameter $A$ is completely determined by the shape parameter $k$. The scale parameter $\lambda$ is irrelevant and it is fixed to $\lambda=1$.  The SD limit is obtained when $k\rightarrow 0$, whereas the WD limit is given by $k\rightarrow \infty$. Indeed, $\lim_{k\rightarrow \infty} f(w) = \delta(w-\lambda)$.

The SD limit is observed at length scales below $\rc$. Thus, when $\rc$ is smaller than the lattice constant, which is assumed to be the unity, the SD-limit effects are completely irrelevant. If we consider now that the numerical prefactor in Eq. \eqref{eq:rc} is of the order of unity, we conclude that $\rc>1$ provided $A>1$. The value of $k$ yielding $A=1$, denoted by $k^\star$, is $k^\star \simeq1.44$ in $D=2$ and $k^\star\simeq 1.16$ in $D=3$. Moreover, $k^\star \rightarrow 1$ when $D\rightarrow \infty$. Therefore, SD-limit effects can only be observed for $k< 1$. We display in Table \ref{table:parameters} the values of $\Anu$ obtained using  Eq. \eqref{eq:A} for some representative values of the shape parameter $k$.

\medskip

\begin{table} [H]
\begin{center}
\begin{tabular}{lccccccccc}
  \hline     \hline
   \backslashbox{$D$}{$k$} & 1 & 0.5 & 0.2 & 0.15 & 0.08 & 0.05 & 0.03 & 0.01 & 0.005\\
   \hline
  $2$ & 1.63 & 4.11  & 13.9 & 20.5 & 47.3 & 88.5 & 174.9 & 756.7 & 1906.7
   \\
  $3$ & 1.14 & 2.09  & 4.69 & 6.04 & 10.5  & 15.9 & 24.9 & 65.5 & 120.5
   \\
  \hline     \hline
\end{tabular}
\caption{Values of $A^\nu$  in $D=2$ and $D=3$ obtained using Eq. \eqref{eq:A} for several values of the shape parameter $k$ of the Weibull distribution. The value of the percolation correlation length exponent is $\nu=4/3$ in $D=2$ and $\nu\approx0.88$ in $D=3$. }
\label{table:parameters}
\end{center}
\end{table}


\section{Scaling ansatz for the mean length of the optimal path}
\label{sec:scaling_ansatz}

A comprehensive numerical study of the model, along with some theoretical arguments based on the results found in the literature, lead us to postulate a unified scaling for the mean length of the optimal path between two points separated a distance $r$ in a disordered $D-$dimensional lattice with lateral linear sizes $\{L_i\}$ ($i=1,\dots,D$). The disorder strength is given by $A$. For the sake of clarity and understanding, we have considered it more appropriate to start by presenting our ansatz and discuss all the scaling regimes derived from it. Next we will show the numerical results that support it.

Our unified scaling ansatz for the mean optimal path length $\loptmean$ has the form
\begin{multline}
  \loptmean\left(r,\{L_i\},A\right) \sim\\
A^\kappa \mathcal{F}\left(\frac{r}{\rc}\right) \mathcal{G}\left(\frac{L}{\Lsat}\right)\mathcal{H}\left(\frac{L}{\min\{r,\rc\}}\right),
\label{eq:general_scaling}
\end{multline}
with the following scaling functions:
\beq
\label{eq:scaling_functions}
\begin{aligned}
 \mathcal{F}(x)&\sim \begin{cases}
                      x^{\alpha} & \mbox{if } x\ll 1, \\
                      x & \mbox{if } x\gg 1,
                    \end{cases} \\
 \mathcal{G}(x)&\sim \begin{cases}
                      x^{\beta} & \mbox{if } x\ll 1, \\
                      1 & \mbox{if } x\gg 1,
                    \end{cases} \\
 \mathcal{H}(x)&\sim \begin{cases}
                      x^{\alpha-1} & \mbox{if } x\ll 1, \\
                      1 & \mbox{if } x\gg 1.
                    \end{cases}
 \end{aligned}
\eeq

Although this ansatz will be analyzed in depth in the following sections, we can highlight now some fundamental aspects for its understanding.

We first note that the initial dependence on the lattice geometry through the set of linear sizes $\{L_i\}$, simplifies to a new variable $L$, defined as the minimum of the lateral sizes of the lattice,
\beq
 L\equiv \min_i\{L_i\}
 \label{eq:L}.
 \eeq
It thus plays the role of a ``reduced'' linear size, and the problem happens to have three degrees of freedom: the two length scales given by $r$ and $L$, and the disorder parameter $A$.

With regard to the three scaling functions, without being too rigorous, we can say that the first scaling function $\mathcal{F}(x)$ controls the scaling of $\loptmean$ with $r$ in strong and weak disorder for fixed $L$. The SD-WD crossover value of $r$, $\rc$, was given in Eq. \eqref{eq:rc}.

The second scaling function, $\mathcal{G}(x)$, do the same but for $L$ with fixed $r$. As we will see, the optimal path also undergoes a disorder-induced transition with respect to $L$, though it is not a crossover. The growth of $\loptmean$ with $L$ saturates when $L$ crosses over a certain saturation value denoted as $\Lsat$ (note that $\mathcal{G}(x)\sim 1$ for $x\gg 1$). For $L\ll \Lsat$, the optimal path behaves (with respect to $L$) as in the SD limit, whereas for $L\gg \Lsat$ the effects of the lattice geometry are irrelevant, as expected for weak disorder. The numerical results show that $\Lsat$ scales like $\rc$,
\beq
\Lsat \sim \Anu,
\eeq
with the numerical prefactors in both scalings being very similar, i.e., $\Lsat \approx \rc$.

The third scaling function $\mathcal{H}(x)$ has a geometrical origin and provides a correction to the previous two scaling functions when $L$ becomes the smallest length scale of the problem.

Finally, the term $A^\kappa$ in Eq. \eqref{eq:general_scaling} stands for the scaling of $\loptmean$ at both the crossover ($r=\rc$) and the saturation ($L=\Lsat$) points.

As will be readily shown, the scaling exponents appearing in Eqs. \eqref{eq:general_scaling} and \eqref{eq:scaling_functions} satisfy the following scaling relation:
\beq
\label{eq:scaling_relation}
\kappa =\nu(\alpha+\beta).
\eeq
Furthermore, we have also found two additional scaling relations between the exponents of our ansatz and the two universal exponents characterizing the behavior of $\loptmean$ in the SD limit. For the fractal dimension $\dopt$ we have
\beq
\label{eq:scaling_relation_dopt}
\dopt=\alpha+\beta,
\eeq
whereas for the universal exponent $-\gopt$ of the power-law decay of the distribution of the optimal path length \cite{35}, we have
\beq
\label{eq:scaling_relation_gopt}
\gopt=1+\frac{\alpha}{\alpha+\beta}=1+\frac{\alpha}{\dopt}.
\eeq

The values of the novel scaling exponents $\alpha$, $\beta$ and $\kappa$, as well as the values of $\nu$, $\dopt$ and $\gopt$ considered here, are indicated in Table \ref{table:scaling_exponents} for $D=2$ and $D=3$. It is important to note that the values of $\alpha$, $\beta$ and $\kappa$ should not be taken as exact values, but as reliable estimates of the actual values. These values were calculated on the basis of two criteria: (i) they provided the best fits to the scaling behaviours obtained in our numerical simulations; (ii) they exactly satisfied the scaling relations given in Eqs.  \eqref{eq:scaling_relation}-\eqref{eq:scaling_relation_gopt}. In this regard, we recall that our aim is to provide a general and consistent theory.

\begin{table}
\begin{center}
\begin{tabular}{ccccccc}
    \hline \hline
   $D$ & $\alpha$ & $\beta$ & $\kappa$ & $\nu$ & $\dopt$ & $\gopt$ \\
  \hline
  $2$ & 0.67 & 0.55 & 1.63 & 4/3 & 1.22 & 1.55  \\
  $3$ & 0.53 & 0.89 & 1.25 & 0.88 & 1.42 & 1.37 \\
  \hline  \hline
\end{tabular}
\caption{Values of the scaling exponents in $D=2$ and $D=3$. The values of $\alpha$, $\beta$ and $\kappa$, were deduced in our work; $\nu$ is a critical exponent of the percolation theory  \cite{46}; the values of $\gopt$ were taken from Ref. \cite{35}; the values of $\dopt$ are the averages of the values reported in several works \cite{20,34,36}.}
\label{table:scaling_exponents}
\end{center}
\end{table}

\section{Scaling regimes}
\label{sec:scaling-regimes}

\begin{figure}
\begin{center}
  \includegraphics[width=\columnwidth]{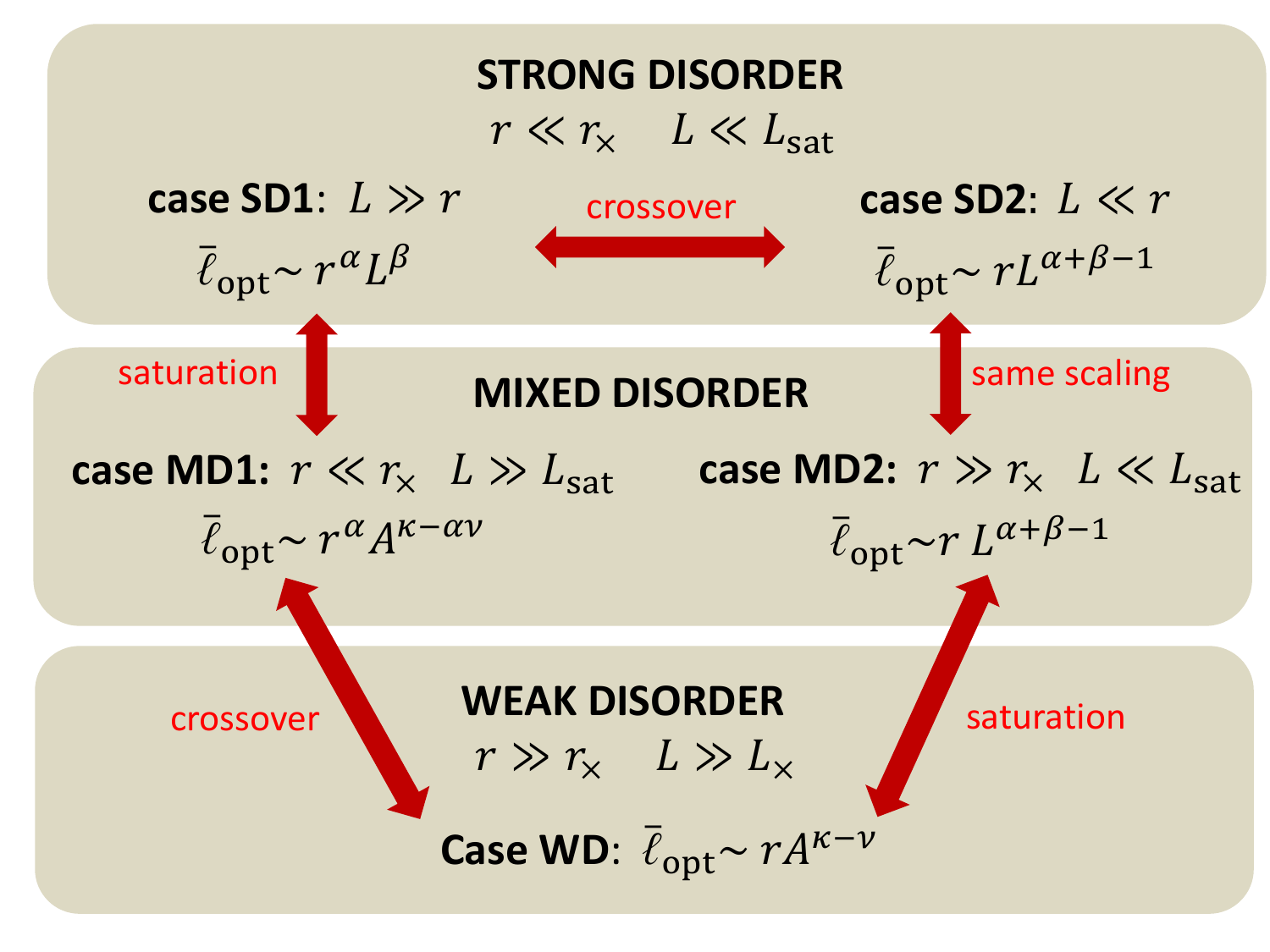}
  \caption{Scheme of the different types of disorder with the five scaling regimes addressed here. The transitions between them have been represented with arrows and the type of transition (crossover or saturation) has been indicated.}
  \label{fig:diagrama_scalings}
\end{center}
\end{figure}

The scaling regimes obtained in our model have been schematically summarized in Fig. \ref{fig:diagrama_scalings}, with the transitions between them indicated by the arrows. We now proceed to their analysis.

\subsection{Strong disorder}

Strong disorder is obtained when $r\ll\rc$ and $L\ll \Lsat$. Under these conditions we
have two possible cases that are addressed separately: $L\gg r$ and $L\ll r$.

\subsubsection{Case SD1: $L \gg r$}

\begin{figure}
\begin{center}
  \includegraphics[width=\columnwidth]{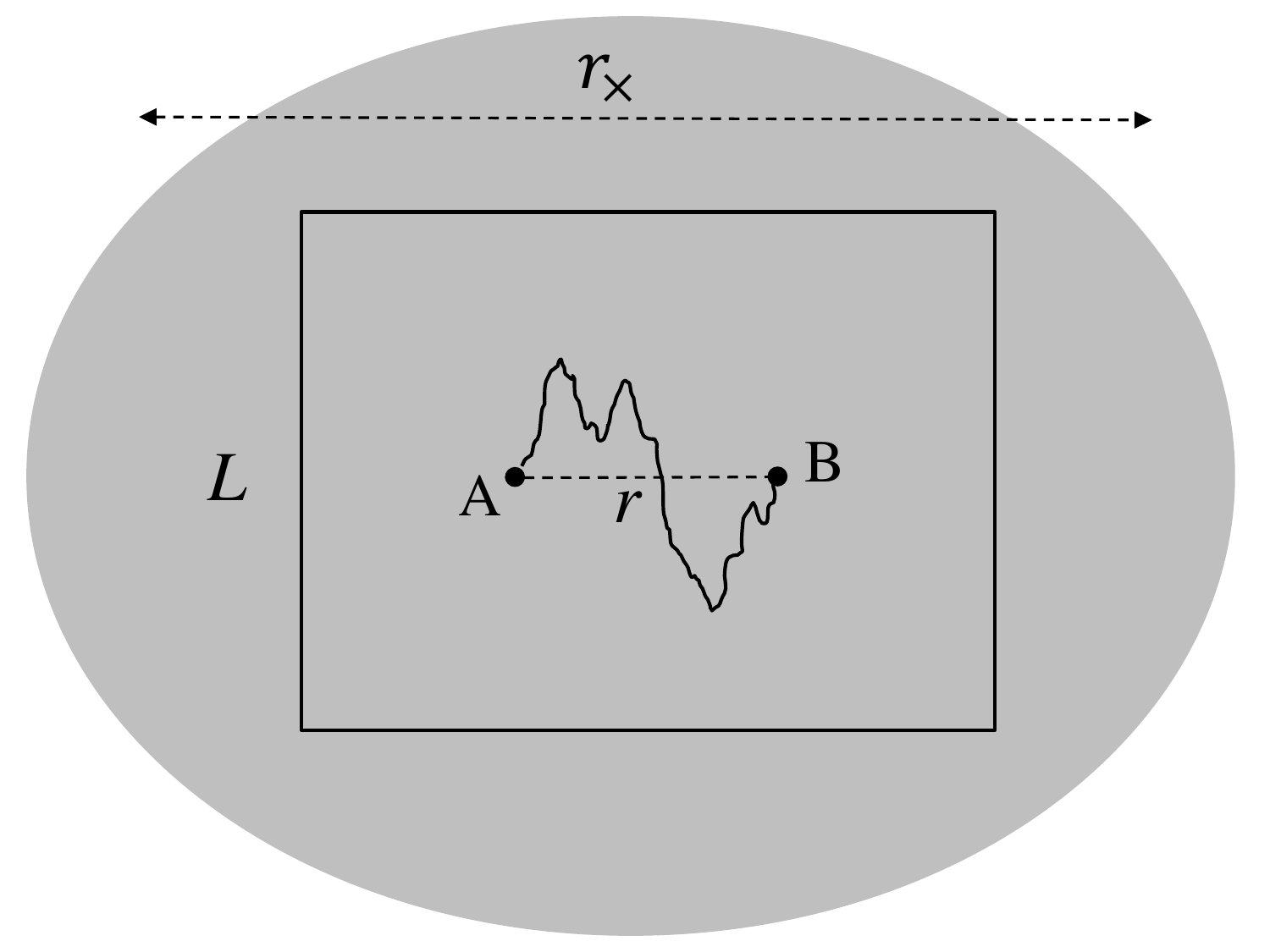}
  \caption{Schematic picture of case SD1 in $D=2$. The lattice is represented by a rectangle where $L$ is the smallest lateral size. The rough line stands for the optimal path between points A and B separated a distance $r$. The gray region represents the domain in which the conditions for the SD limit are satisfied, and has a linear size of $\rc$ (we recall that $\Lsat \approx \rc$).}
    \label{fig:esquema_SD1}
\end{center}
\end{figure}

Figure \ref{fig:esquema_SD1} illustrates the geometric conditions corresponding to this case. The lattice is under SD-limit conditions and the reduced linear size $L$ is larger than $r$.

From the general ansatz presented in Eq. \eqref{eq:general_scaling} we deduce the following scaling, called \emph{case SD1}:
\beq
\label{eq:SD1}
\loptmean \sim r^\alpha L^{\beta},  \qquad (\text{SD1})
\eeq
where we have made use of the scaling relation in Eq. $\eqref{eq:scaling_relation}$. Since the optimal path evolves in the SD limit, the scaling does not depend on the disorder strength $A$. We compare now this scaling to known results.

The probability density of the optimal path length, $P(\lopt|r,L)$, in square and simple cubic lattices of linear size $L\gg r$, was deduced in Ref. \cite{35} from numerical simulations. It was shown that
\begin{multline}
  P(\lopt|r,L)\sim\\
\frac{1}{r^{\dopt}} \left[\frac{\lopt}{r^{\dopt}}\right]^{-{\gopt}}
f_1\left(\frac{\lopt}{r^{\dopt}}\right)f_2\left(\frac{\lopt}{L^{\dopt}}\right).
\label{eq:pdf_lopt}
\end{multline}

Functions $f_1$ and $f_2$ were assumed to be stretched exponential functions that model the steep drop of $P(\lopt|r,L)$ near certain lower and upper cutoffs for the optimal path length. We denote these bounds as $\loptmin$ and $\loptmax$ respectively.

The lower cutoff value $\loptmin$ is related to the most probable value $\loptstar$, and is expected to scale as the mean optimal path length \cite{35}, $\loptmin\approx \loptstar \sim r^{\dopt}$. It appears to be independent of $L$ as long as $L\gg r$. The upper cutoff value $\loptmax$ is due to the finite size $L$, and seems to scale as $\loptmax \sim L^{\dopt}$. It appears to be independent of $r$ as long as $L\gg r$.

For $\loptmin \ll \lopt \ll \loptmax $, $P(\lopt|r,L)$ shows a power-law decay with exponent $-\gopt\simeq-1.55$ in $D=2$, and $-\gopt\simeq -1.37$ in $D=3$. Note that $\gopt$ decreases as $D$ increases, which means that the longer optimal paths have a larger probability at larger dimensions. Additionally, since $\gopt<2$ for all $d$, $\loptmean$ and all higher moments diverge as $L\rightarrow \infty$.

We simplify now Eq. \eqref{eq:pdf_lopt} by replacing the stretched exponential functions by unit step functions, and assume that all the numerical prefactors in the scaling laws are of the order of unity. We then have
\begin{multline}
  P(\lopt|r,L)=\\
\begin{cases}
               \frac{1}{N} \left[\frac{\lopt}{r^{\dopt}}\right]^{-{\gopt}} & \mbox{if} \quad \lopt\in[r^{\dopt},L^{\dopt}], \\
               0 & \mbox{otherwise},
             \end{cases}
\label{eq:pdf_lopt_simpli}
\end{multline}
where $N$ is a normalization prefactor. The mean optimal path length of that distribution is
\beq
\label{eq:loptmean_dist}
\loptmean(r,L)=\frac{(\gopt-1)}{(2-\gopt)}
\frac{(r^\alpha L^{\dopt} -r^{\dopt}L^{\alpha})}{(L^\alpha-r^\alpha)},
\eeq
where $\alpha={\dopt}({\gopt}-1)$. In we apply now condition $L\gg r$, we obtain $\loptmean\sim r^\alpha L^{{\dopt}-\alpha}$, which is the same as the scaling of case SD1 given in Eq. \eqref{eq:SD1} provided $\beta=\dopt-\alpha$. This leads to the scaling relations given in Eqs. \eqref{eq:scaling_relation_dopt} and \eqref{eq:scaling_relation_gopt}.

Let us consider now an isotropic scaling of the two linear scales, $L\sim r$, which has been by far the most-addressed case \cite{11,32,33,36}. If we apply this condition to the SD1 scaling given in Eq. \eqref{eq:SD1}, we obtain  $\loptmean \sim r^{\alpha+\beta}= r^{\dopt}$. Therefore, the so-called \emph{universal} --or \emph{fractal}-- \emph{scaling} in the SD limit arises naturally from our model when the system is scaled isotropically, as expected.

\subsubsection{Case SD2: $L\ll r$}

\begin{figure}
\begin{center}
  \includegraphics[width=\columnwidth]{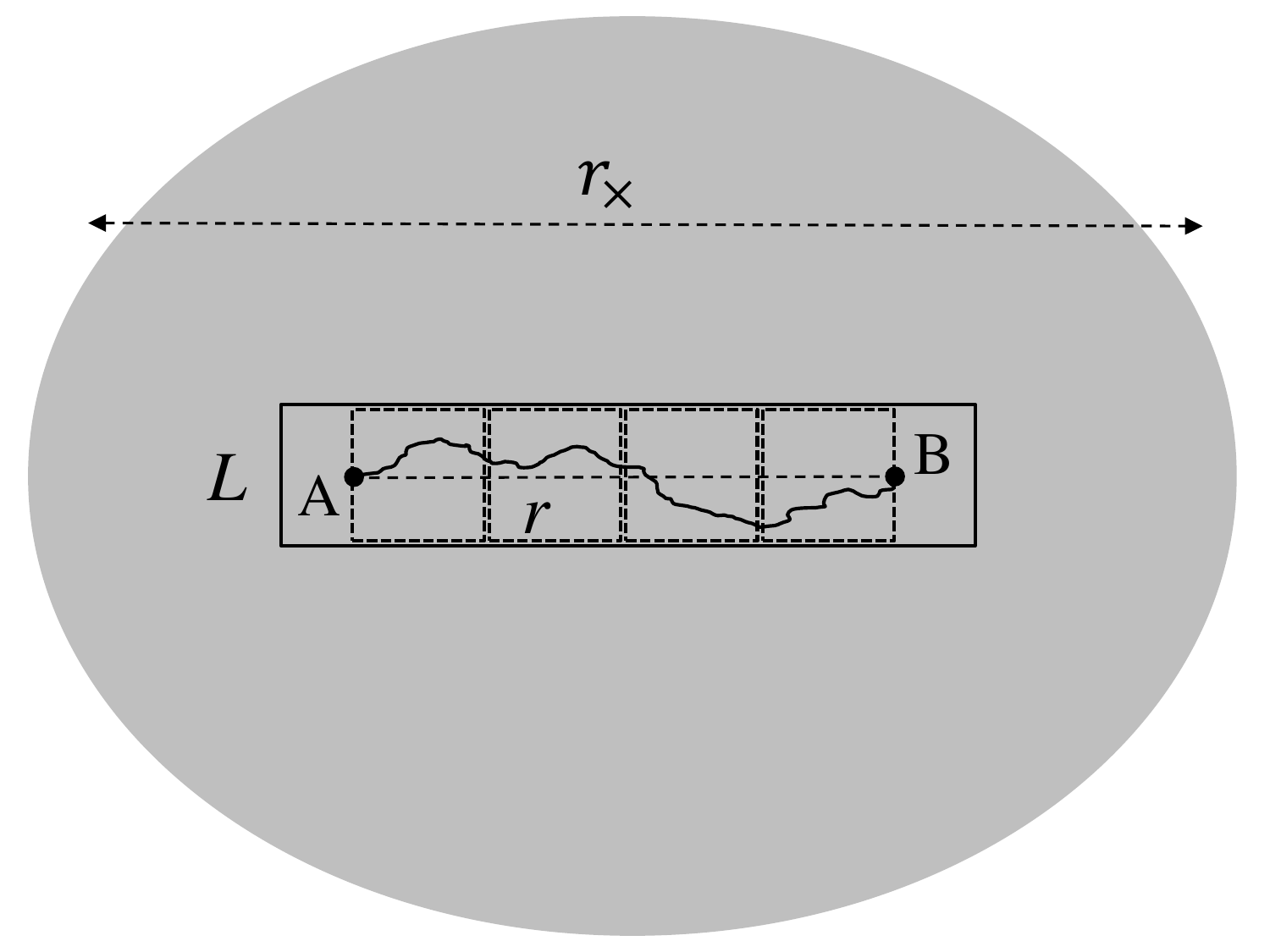}
  \caption{Schematic picture of case SD2 in $D=2$. The lattice is represented by a rectangle where $L$ is the smallest lateral size. The rough line stands for the optimal path between points A and B separated a distance $r$. The gray region represents the domain in which the conditions for the SD limit are satisfied, and has a linear size of $\rc$ (we recall that $\Lsat \approx \rc$). The squares with dotted line represent patches of linear size $L$. }
    \label{fig:esquema_SD2}
\end{center}
\end{figure}

Figure \ref{fig:esquema_SD2} illustrates the geometric conditions corresponding to this case. The lattice is in the SD limit and $L$ is smaller than $r$.

The general scaling in Eq. \eqref{eq:general_scaling} yields a novel scaling regime called \emph{case SD2},
  \beq
  \label{eq:SD2}
  \loptmean \sim rL^{\alpha + \beta-1}=rL^{\dopt-1}  \qquad (\text{SD2}).
  \eeq

As illustrated in Fig. \ref{fig:esquema_SD2}, the optimal path can be divided into $r/L$ patches of linear size $L$. Within each patch, $\loptmean$ behaves as in case SD1 with $r=L$, $\loptmean \sim L^{\dopt}$. We then have $\loptmean \sim (r/L)L^{\dopt}$ hence obtaining the above scaling.

\subsection{Mixed disorder}

We use the term \emph{mixed disorder} to refer to cases where the disorder is strong with respect to one of the linear scales of the model, but it is weak with respect to the other. Since we have two linear scales, $r$ and $L$, we obtain two different cases (see Fig. \ref{fig:diagrama_scalings}).

\subsubsection{Case MD1: $r\ll\rc$ and $L\gg\Lsat$}

\begin{figure}
\begin{center}
  \includegraphics[width=\columnwidth]{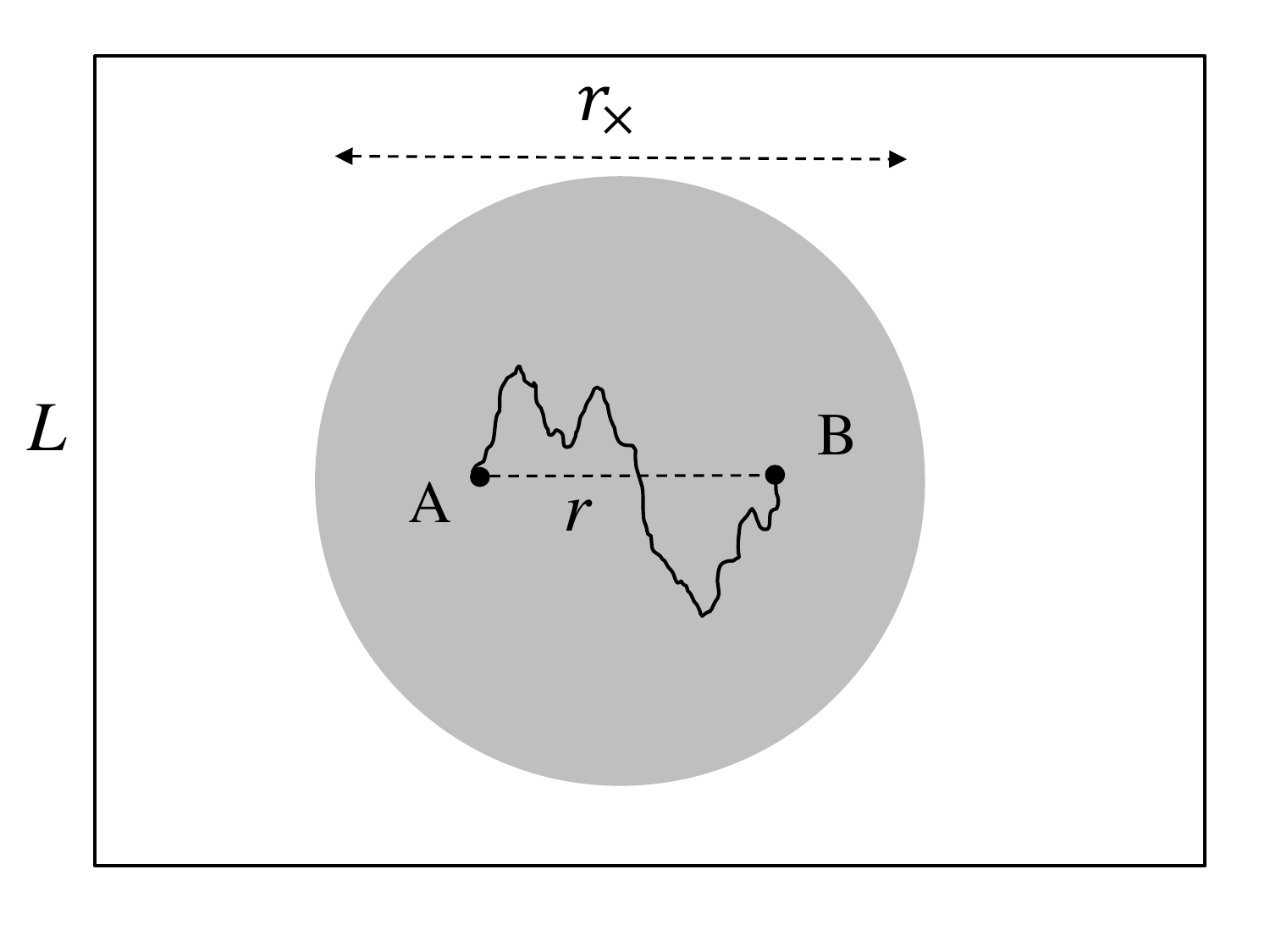}
  \caption{Schematic picture of case MD1 in $D=2$. The lattice is represented by a rectangle where $L$ is the smallest lateral size. The rough line stands for the optimal path between points A and B separated a distance $r$. The gray region represents the domain in which the conditions for the SD limit are satisfied, and has a linear size of $\rc$ (we recall that $\Lsat \approx \rc$).  }
    \label{fig:esquema_MD1}
\end{center}
\end{figure}

Figure \ref{fig:esquema_MD1} illustrates the geometric conditions corresponding to this case. The optimal path behaves with respect to $r$ as in the SD limit, but the effects of the lattice geometry are completely irrelevant. Notice that $L\gg r$, so this case accounts for optimal paths in very large lattices compared to $r$. From Eq. \eqref{eq:general_scaling} we obtain the following scaling, called \emph{case MD1}:
  \beq
    \label{eq:MD1}
  \loptmean \sim r^\alpha A^{\kappa-\alpha\nu}= r^\alpha A^{\nu(\dopt -\alpha)}   \qquad (\text{MD1}).
  \eeq
As expected, the optimal path length does not vary with $L$ because it saturates at $\Lsat$. Accordingly, scaling MD1 is obtained by making $L=\Lsat\sim A^\nu$ in case SD1. For a given disorder strength $A$, the optimal path length scales with the end-to-end distance $r$ with exponent $\alpha$.

\subsubsection{Case MD2: $r\gg\rc$ and $L\ll\Lsat$}

\begin{figure}
\begin{center}
  \includegraphics[width=\columnwidth]{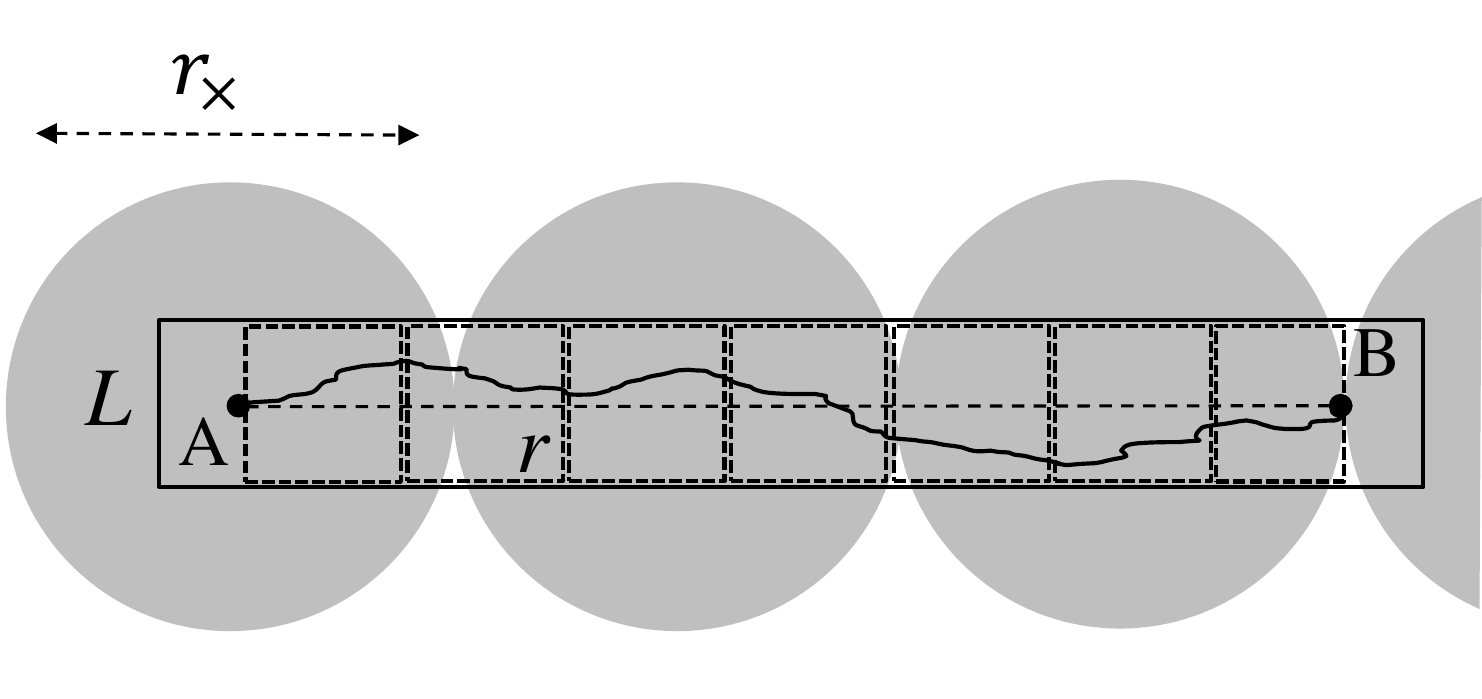}
  \caption{Schematic picture of case MD2 in $D=2$. The lattice is represented by a rectangle where $L$ is the smallest lateral size. The rough line stands for the optimal path between points A and B separated a distance $r$. The gray region represents the domain in which the conditions for the SD limit are satisfied, and has a linear size of $\rc$ (we recall that $\Lsat \approx \rc$). The squares with dotted line represent patches of linear size $L$.}
    \label{fig:esquema_MD2}
\end{center}
\end{figure}

In this case we necessary have $L\ll r$. Figure \ref{fig:esquema_MD2} illustrates the geometric conditions applying to this case. The ``effective'' lattice of linear size $L$ is in the SD limit, but the end-to-end distance is much larger than $\rc$, so the optimal path ``sees'' a weak disorder along this direction.

From Eq. \eqref{eq:general_scaling} we obtain the scaling called \emph{case MD2},
  \beq
      \label{eq:MD2}
  \loptmean \sim r L^{\alpha + \beta-1}=r L^{\dopt-1} \qquad (\text{MD2}),
  \eeq
which is equal to scaling SD2, so the same arguments employed for that case also apply here.

\subsection{Weak disorder}

\begin{figure}
\begin{center}
  \includegraphics[width=\columnwidth]{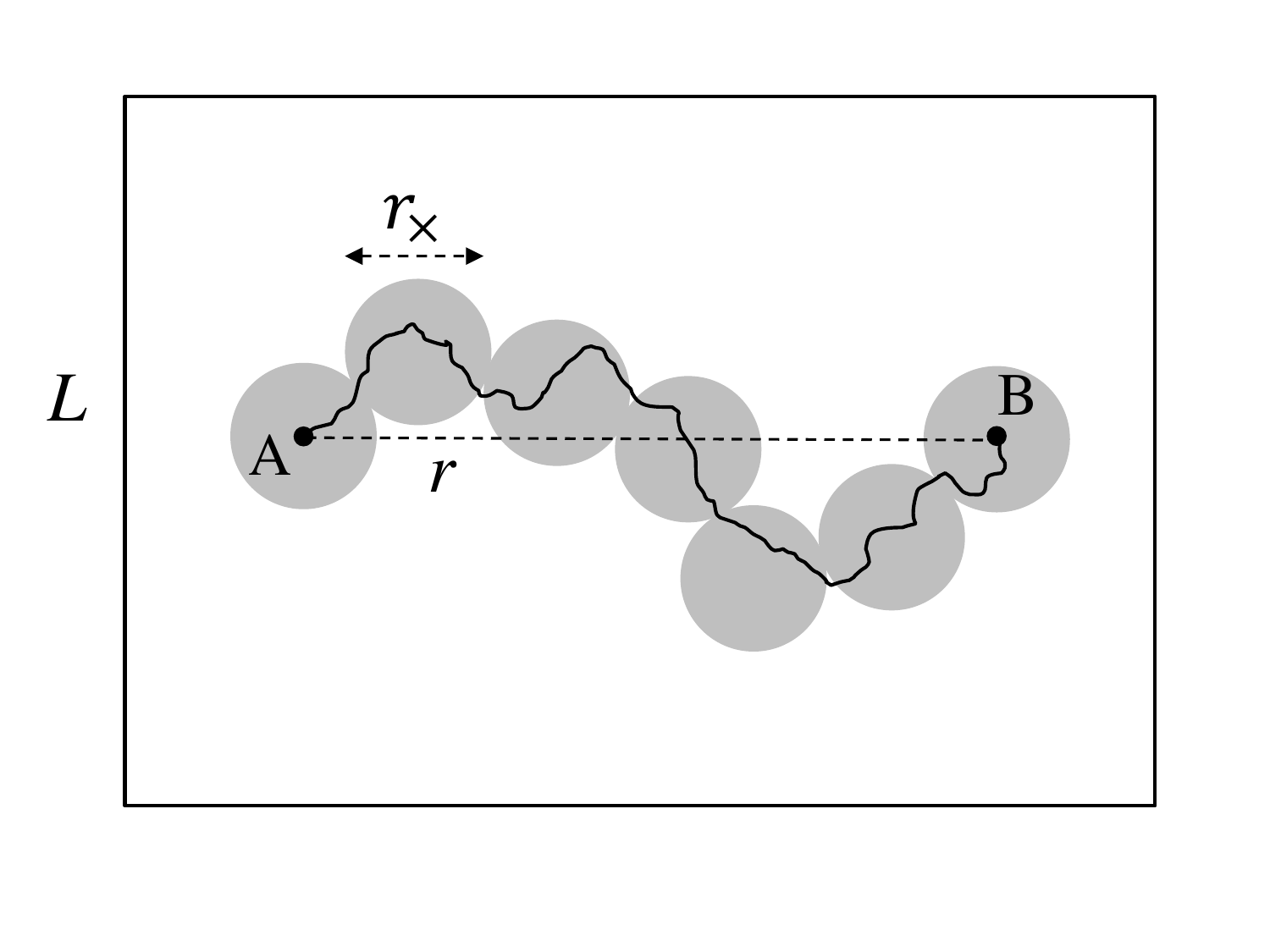}
  \caption{Schematic picture of case WD in $D=2$. The lattice is represented by a rectangle where $L$ is the smallest lateral size. The rough line stands for the optimal path between points A and B separated a distance $r$. The gray region represents the domain in which the conditions for the SD limit are satisfied, and has a linear size of $\rc$ (we recall that $\Lsat \approx \rc$).}
    \label{fig:esquema_WD}
\end{center}
\end{figure}

Weak disorder is obtained when $r\gg \rc$ and $L\gg \Lsat$. Figure \ref{fig:esquema_WD} illustrates the geometric conditions corresponding to this case. The relation between $r$ and $L$ is irrelevant as long as the above conditions are satisfied.

From Eq. \eqref{eq:general_scaling} we obtain \emph{case WD},
  \beq
       \label{eq:WD}
  \loptmean \sim rA^{\kappa-\nu}= rA^{\nu(\dopt-1)}     \qquad (\text{WD}),
  \eeq
which coincides with the WD scaling reported in the literature \cite{23,32,33,34}.

\section{Discussion}
\label{sec:discussion}

Transitions between the five scaling regimes described above, indicated in Fig. \ref{fig:diagrama_scalings} by arrows, are distinguished into two classes: saturations and crossovers.

Saturations take place when the effective lattice size $L$ crosses over the saturation value $\Lsat$, and are mediated by the scaling function $\mathcal{G}(x)$ in Eq. \eqref{eq:scaling_functions}. Saturations are observed in the transitions SD1$\shortrightarrow$MD1 and MD2$\shortrightarrow$WD. To obtain the final scaling we only need to replace $L$ by $\Anu$ in the initial scaling.

Crossovers mean changes of the scaling exponents of $r$ and $L$. The first crossover occurs in the SD limit between SD1 and SD2, and takes place when $L$ becomes of the order of $r$. It is mediated by the scaling function $\mathcal{H}(x)$ in Eq. \eqref{eq:scaling_functions}. The crossover value of $\loptmean$, denoted as $\loptmeancA$, is obtained at $L=r$ and scales as $\loptmeancA \sim r^{\dopt}$. It thus stands for the fractal scaling reported in the literature \cite{11,32,33,36}.

The second crossover is the transition between MD1 and WD, and takes place when the lattice size effects are irrelevant ($L\gg \Lsat$) and the en-to-end distance $r$ reaches its crossover value $\rc$. The crossover mean length, denoted as $\loptmeancB$, scales with disorder as
\beq
\loptmeancB \sim A^{\kappa}=A^{\nu\dopt},
\label{eq:crossover_lenght}
\eeq
which is in perfect agreement with previous results \cite{17,17b,23,32,33}.

A specially interesting case of the second crossover happens when $L=r$. Then, the crossover takes place directly between SD1 and WD because both $r$ and $L$ cross over their transition values $\rc$ and $\Lsat$, respectively, at the same time. This is the transition that has been recognized in the literature as the strong-weak disorder crossover \cite{23,32,33,34}.

Finally, the WD scaling presented in Eq. \eqref{eq:WD} can be interpreted as follows. As illustrated in Fig. \ref{fig:esquema_WD}, for WD conditions the optimal path becomes equivalent to a directed polymer in an effective lattice consisting of cells of linear size $\rc$ \cite{33}. The number of cells scales as $r$ because the  transversal deviation of the optimal path, $h$, scales with $r$ with an exponent smaller than 1. Indeed, $h\sim r^{\zeta}$,  where $\zeta$ is the DPRM wandering exponent ($\zeta=2/3$ in $D=2$ and $\zeta\simeq 0.62$ in $D=3$ \cite{8,50}), which is related to the KPZ dynamical exponent $z$ through $\zeta=1/z$ \cite{8,14,50}. The length of the optimal path inside each cell is $\loptmeancB \sim A^{\kappa}$. Thus, we can write $\loptmean \sim (r/\rc)\loptmeancB \sim rA^{\kappa-\nu}$, and we obtain the WD scaling given in Eq. \eqref{eq:WD}.

We would like to end our discussion by addressing a disagreement that has gone completely unnoticed. Porto et al. \cite{34} obtained that $\loptmeancB$ scales with disorder according to Eq. \eqref{eq:crossover_lenght}, but with an exponent $\kappa \simeq 1.60$ in both $D=2$ and $D=3$.  While the value in $D=2$ is in good agreement with $\kappa =\nu\dopt\simeq 1.63$, the value in $D=3$ is considerable different, $\kappa =\nu\dopt \simeq 1.25$. In that work, the authors studied numerically the scaling of the radius of gyration of the optimal path, denoted here by $R_g$, with the path length $\lopt$, in square and simple cubic lattices with fixed size $L$ and for different disorder strengths. 

These results offer us the possibility of testing the ability of our model to deduce the behavior of other observables such as the radius of gyration. If we analyze the conditions used in the work, we find that $L\gg \Lsat$ for all noise levels considered. According to our theory, the strong-weak disorder crossover shown in the work actually corresponds to the crossover from case MD1 to WD. For a fractal polymer of length $\ell$ and fractal dimension $d_f$, its radius of gyration scales as $R_g \sim \ell^{1/d_f}$. This is equivalent to the fractal law $m\sim R^{d_f}$ for the mass $m$ of a fractal that is contained in a hypersphere of radius $R$. Therefore, $R_g$ scales as the linear size $R$ of the hypervolume containing the polymer. If that volume is a $D$-dimensional lattice of size $L_1\times L_2\times \ldots\times L_D$, we then expect $R_g\sim (L_1 L_2\cdots L_D)^{1/D}$. In the MD1 case we have the configuration illustrated in Fig. \ref{fig:esquema_MD1}. Since $L\gg \Lsat \gg r$, the optimal path behaves as if the lateral size of the lattice in all directions except for the $r$-direction was given by $\Lsat$. Thus, the average linear size $R$ of the volume comprised by the optimal path should scale as $R\sim \left(rA^{\nu(D-1)}\right)^{1/D}$, so we expect for the MD1 case $R_g\sim \left(rA^{\nu(D-1)}\right)^{1/D}$. On the other hand, in the WD case the optimal path behaves as a self-affine directed polymer, so we expect $R_g\sim r$. We then conclude
\beq
R_g^2 \sim \begin{cases}
           r^{2/D}A^{2\nu(D-1)/D} & \mbox{if } r\ll \rc, \\
           r^2 & \mbox{if } r\gg \rc.
         \end{cases}
         \label{eq:R_g1}
\eeq

Now we use the scaling of $\loptmean$ in cases MD1 ($r\ll \rc$) and WD ($r\gg \rc$), given in Eqs. \eqref{eq:MD1} and \eqref{eq:WD} respectively, to obtain $r$ as a function of $\loptmean$. By introducing the resulting expressions in Eq. \eqref{eq:R_g1} we obtain
\beq
R_g^2 \sim \begin{cases}
           \loptmean^{,2/\alpha D} A^{2\nu(1-\dopt/\alpha D)} & \mbox{if} \quad  \loptmean\ll \loptmeancB, \\
           \loptmean^{,2} A^{2\nu(1-\dopt)} & \mbox{if} \quad \loptmean \gg \loptmeancB.
         \end{cases}
         \label{eq:R_g2}
\eeq
The crossover mean length is the value of $\loptmean$ that makes the two parts equal, and coincides with $\loptmeancB$ given in Eq. \eqref{eq:crossover_lenght}.

We now rewrite the above scaling in the form
\beq
R_g^2 \sim \left(\loptmeancB\right)^{2/\dopt}\mathcal{Z}\left(\frac{\loptmean}{\loptmeancB}\right),
\label{eq:R_g3}
\eeq
with the scaling function
\beq
 \mathcal{Z}(x)\sim \begin{cases}
                      x^{2/\alpha D} & \mbox{if } x\ll 1, \\
                      x^2 & \mbox{if } x\gg 1,
                    \end{cases} \\
\eeq
and we obtain the same scaling reported in that work provided $\alpha=\dopt/D$ (see Fig. 2 in Ref. \cite{34}). In conclusion, our theory successfully reproduces the scaling of the radius of gyration, but disagrees on the value of $\kappa$ in $D=3$. Interestingly, a new scaling relation $\alpha=\dopt/D$ has been deduced. Its validation is beyond the scope of the present work, but it certainly deserves an in-depth study. The values of $\alpha$ obtained from it do not differ notably from those considered here: in $D=2$ we have $0.61$ vs. $0.67$ respectively, and in $D=3$ we have $0.47$ vs. $0.53$.

\section{Numerical results}
\label{sec:num_results}

Now we present a series of numerical experiments which are intended to show the role of the different terms involved in our unified scaling ansatz, as well as to illustrate the afore-discussed scaling regimes and transitions. In particular, we will focus on showing the scaling functions given in Eq. \eqref{eq:scaling_functions}. We recall that we tune the strength of the disorder by varying the shape parameter $k$ of the Weibull distribution, and that disorder strength parameter $A$ is obtained from $k$ using Eq. \eqref{eq:A}. However, in the following figures we will represent the degree of disorder by the term $A^\nu$ because it provides more information as it determines the scaling of both $\rc$ and $\Lsat$. Thus, a given value of $A^\nu$ corresponds to a certain value of $k$ (see e.g., Table \ref{table:parameters} for some representative values). Unless we say the contrary, the averages are performed over $5\times 10^3$ paths.

\begin{figure}
  \includegraphics[width=\columnwidth]{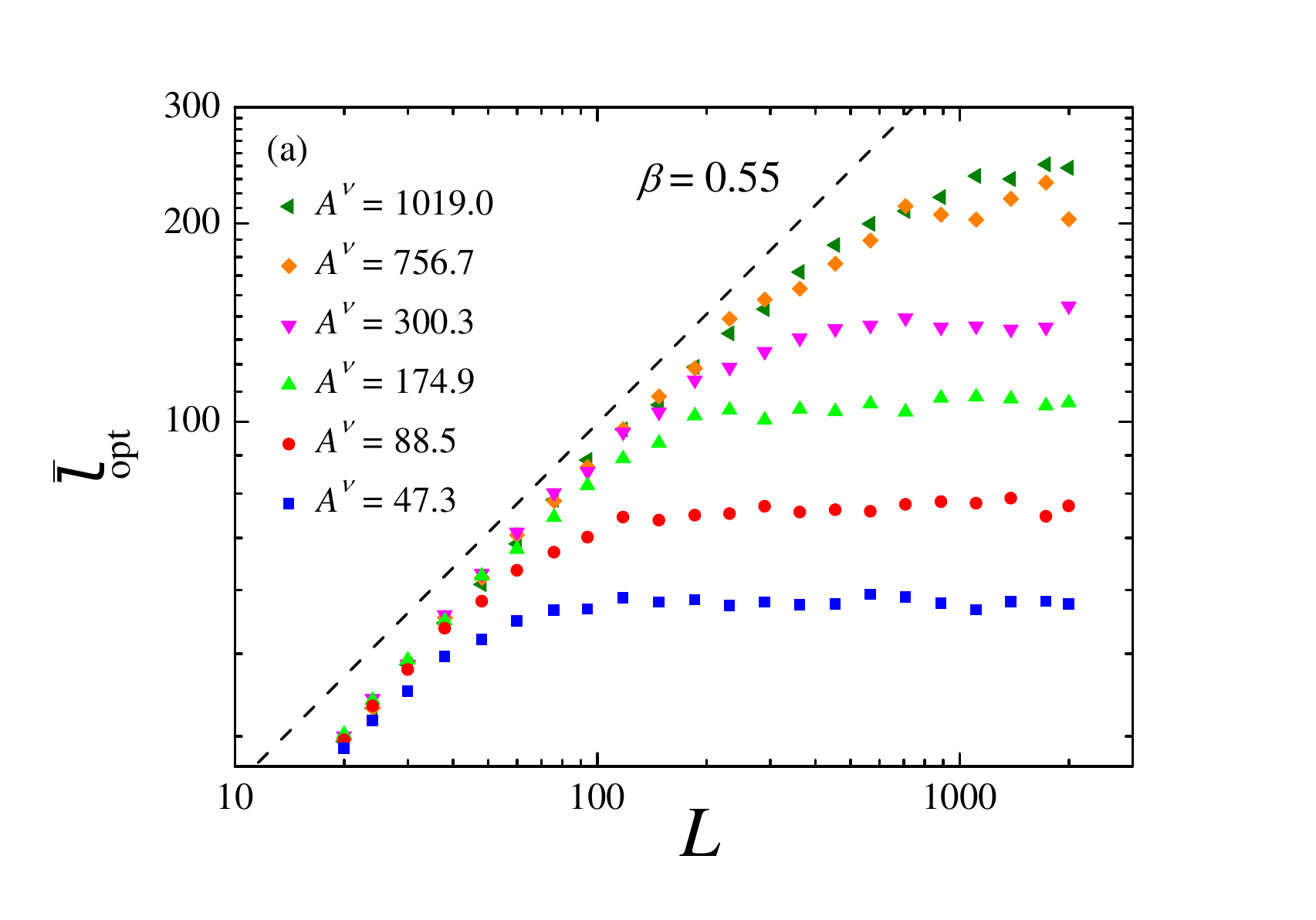}
  \includegraphics[width=\columnwidth]{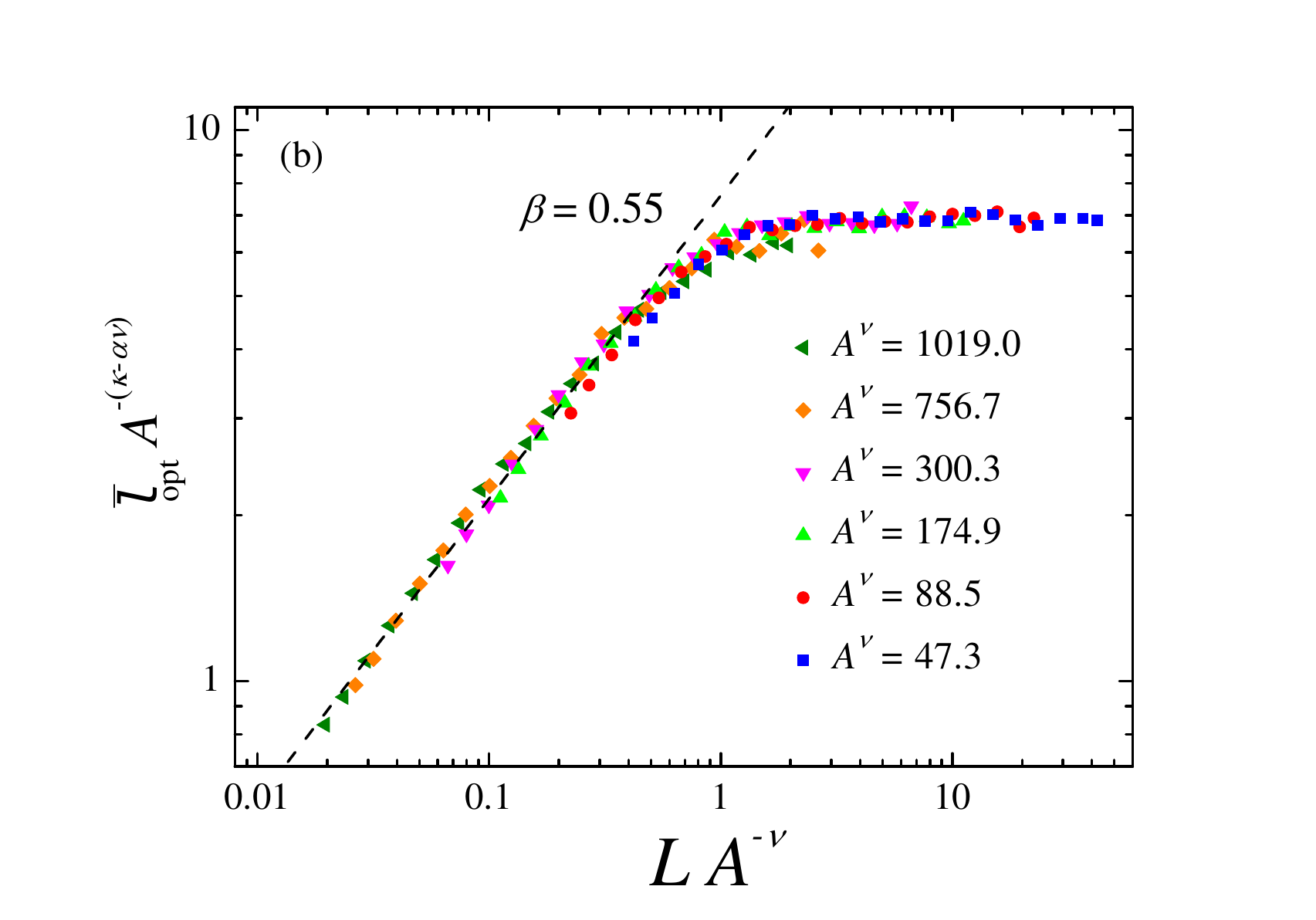}
  \caption{(a) Mean optimal path length $\loptmean$ as a function of the linear size $L$ of square lattices for $r=10$ and different disorder strengths. (b) Data collapse obtained after applying the scaling given in Eq. \eqref{eq:scaling_rfijo} with exponents given in Table \ref{table:scaling_exponents} for $D=2$. Broken lines represent power law behavior with exponent $\beta$.}
  \label{fig:scaling_rfijo}
\end{figure}

One of most interesting results of our study is the saturation of the optimal path length when $L$ crosses over $\Lsat$. Figure \ref{fig:scaling_rfijo}(a) shows the scaling of $\loptmean$ with the linear size $L$ of square lattices, for fixed $r\ll \rc$ and different degrees of disorder. We are thus displaying the transition from SD1 to MD1 with fixed $r$. From the general ansatz given in Eq. \eqref{eq:general_scaling} we obtain the following scaling:
\beq
\label{eq:scaling_rfijo}
\loptmean \sim  A^{\kappa-\alpha\nu} \mathcal{G}\left(\frac{L}{A^\nu}\right),
\eeq
which has the same form as the Family-Vicsek scaling for the roughness of growing surfaces \cite{50}.

The results displayed in the figure are in good agreement with this scaling. The mean optimal path length grows as $L^{\beta}$ up to a saturation size $\Lsat$ which increases with disorder. From this point on, $\loptmean$ saturates at a value that also increases with disorder. In Fig. \ref{fig:scaling_rfijo}(b) we have scaled $L$ by the saturation size $\Lsat\sim A^\nu$, and $\loptmean$ by its saturation value, which scales as $A^{\kappa-\alpha\nu}$. Data nicely collapse to the scaling function $\mathcal G(x)$ given in Eq. \eqref{eq:scaling_functions}. The corresponding Family-Vicsek relation between the exponents takes the form $\kappa=\nu(\alpha+\beta)$ and was previously presented in Eq. \eqref{eq:scaling_relation}.

We focus now on the crossover from MD1 to WD. We fix $L$ to a constant value such that $L \gg \Anu$, and we vary $r$ for different values of $A$. Under these conditions our ansatz in Eq. \eqref{eq:general_scaling} reduces to
\beq
\label{eq:scaling_Lfijo}
\loptmean \sim A^\kappa \mathcal{F}\left(\frac{r}{A^\nu}\right).
\eeq

\begin{figure}
  \includegraphics[width=\columnwidth]{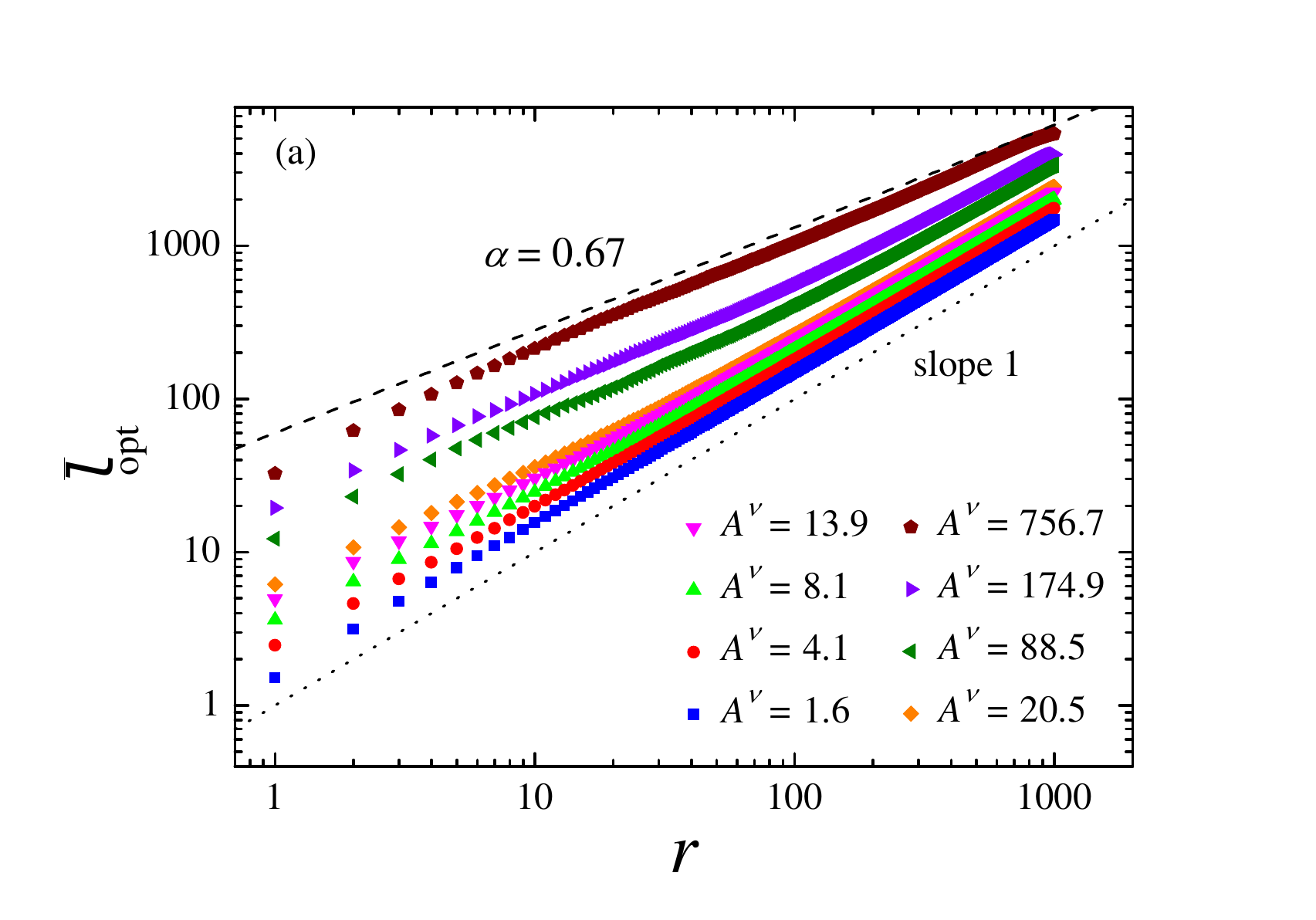}
  \includegraphics[width=\columnwidth]{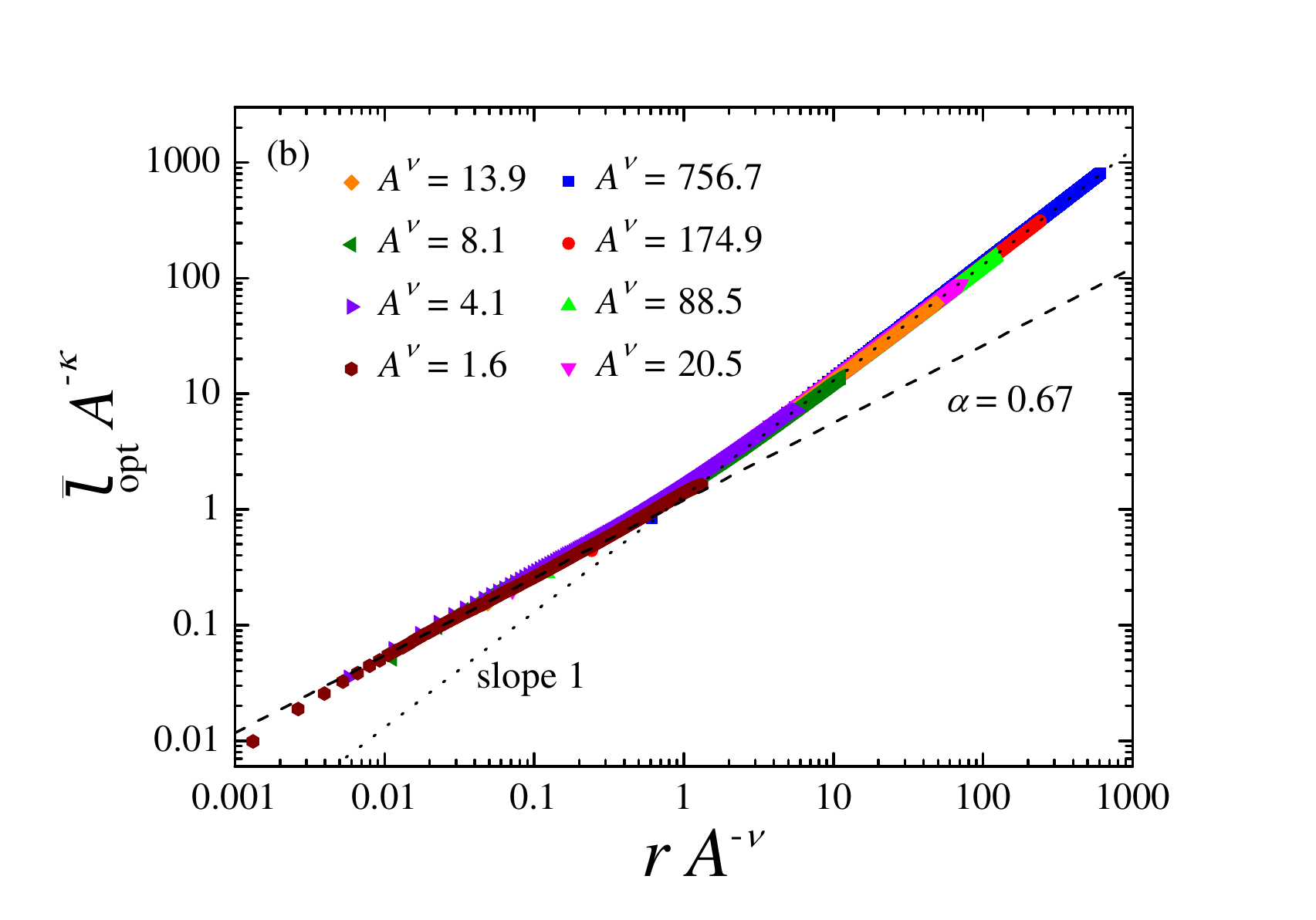}
  \caption{  (a) Mean length of the optimal path connecting the center node of a square lattice of linear size $L=2000$, to nodes on the axis, as a function of their distance $r$ to the origin for different disorder strengths. (b) Data collapse obtained after applying the scaling given in Eq. \eqref{eq:scaling_Lfijo} with exponents given in Table \ref{table:scaling_exponents} for $D=2$. Broken lines in both panels represent power law behaviors with the indicated exponents. The averages are performed over $4\times 10^4$ paths.}
  \label{fig:scaling_Lfijo}
\end{figure}

We have carried out this numerical experiment on square lattices of linear size $L=2000$ with the following particularity. We have considered the optimal path connecting the center node to nodes on the axis, so $r$ is the Euclidean distance to the origin, $r=1,\ldots,1000$. The increase of the average length of these paths with $r$ has been displayed in Fig. \ref{fig:scaling_Lfijo}(a) for several disorders ranging from $\Anu\approx 2$ to $\Anu\approx 760$. In Figure \ref{fig:scaling_Lfijo}(b) we have scaled data according to Eq. \eqref{eq:scaling_Lfijo}. Despite the fact that the optimal path ``sees'' a different lattice geometry as $r$ increases, data nicely collapse to the scaling function $\mathcal{F}(x)$ given in Eq. \eqref{eq:scaling_functions}. Small deviations are observed for the smaller values of $r$, and are due to the discreteness of the lattice, which becomes relevant.

The two set of results discussed above showed the transition from SD1 to WD in two steps: first SD1$\shortrightarrow$MD1 and then MD1$\shortrightarrow$WD. However, as discussed in Sec. \ref{sec:discussion}, we can simplify this path by assuming that $L=r$. In that case the transition from the SD limit to WD takes place in one single step because both $r$ and $L$ cross over $\Anu$ simultaneously. From the general ansatz given in Eq. \eqref{eq:general_scaling} we obtain
\beq
\label{eq:scaling_r=L}
\loptmean\sim A^\kappa \mathcal{J}\left(\frac{r}{\Anu}\right),
\eeq
with scaling function $\mathcal{J}(x)=\mathcal{F}(x)\mathcal{G}(x)$:
\beq
\label{eq:scaling_function_L=r}
 \mathcal{J}(x)\sim \begin{cases}
                      x^{\dopt} & \mbox{if } x\ll 1, \\
                      x & \mbox{if } x\gg 1.
                    \end{cases}
\eeq

This transition has been reported in the literature as the strong-weak disorder crossover \cite{23,32,33,34}. The scaling factor $A^\kappa$ and the scaling function $\mathcal{J}(x)$ may take different forms, but they all lead to the same scaling behaviors on both sides of the crossover.

\begin{figure}
  \includegraphics[width=\columnwidth]{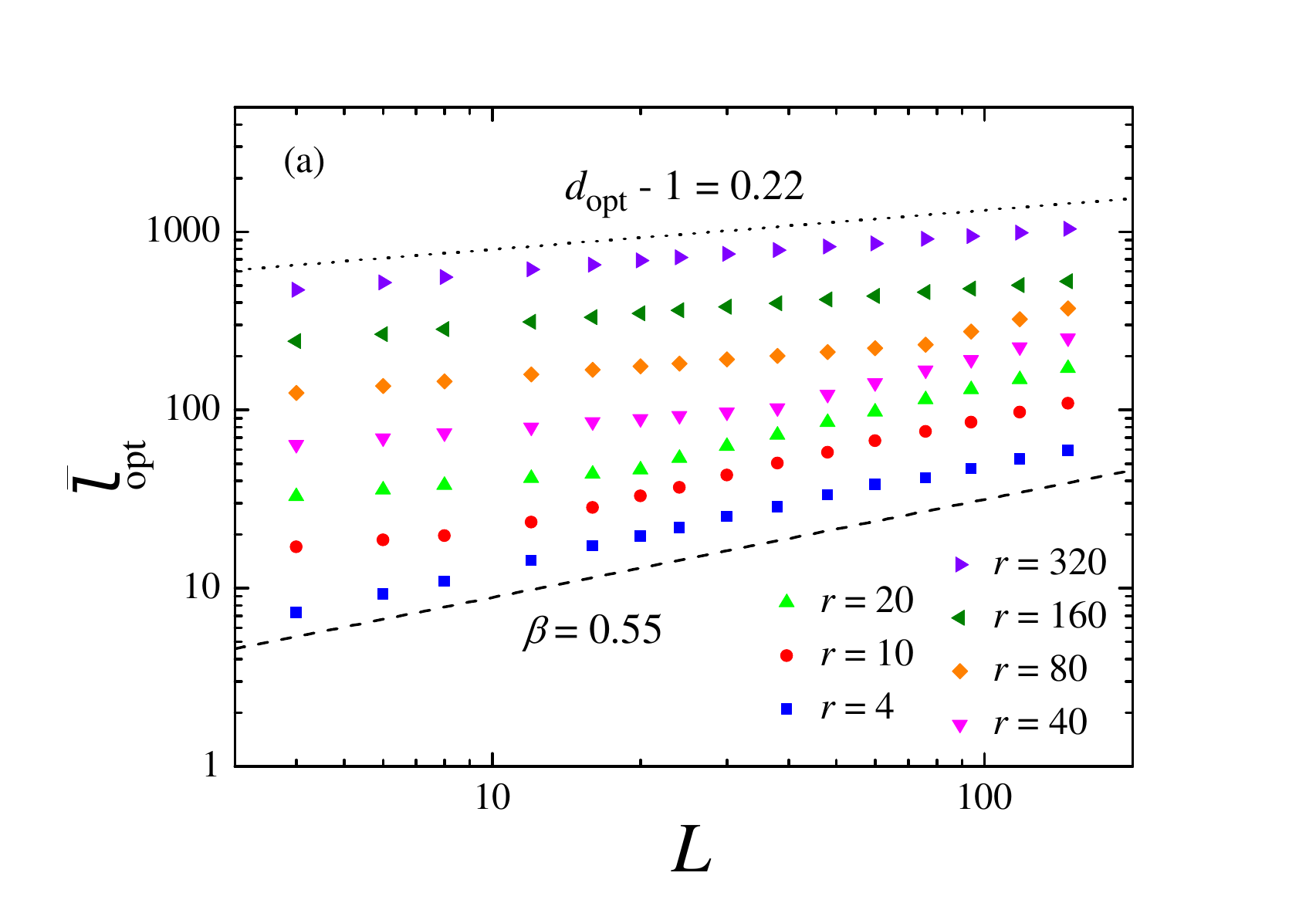}
  \includegraphics[width=\columnwidth]{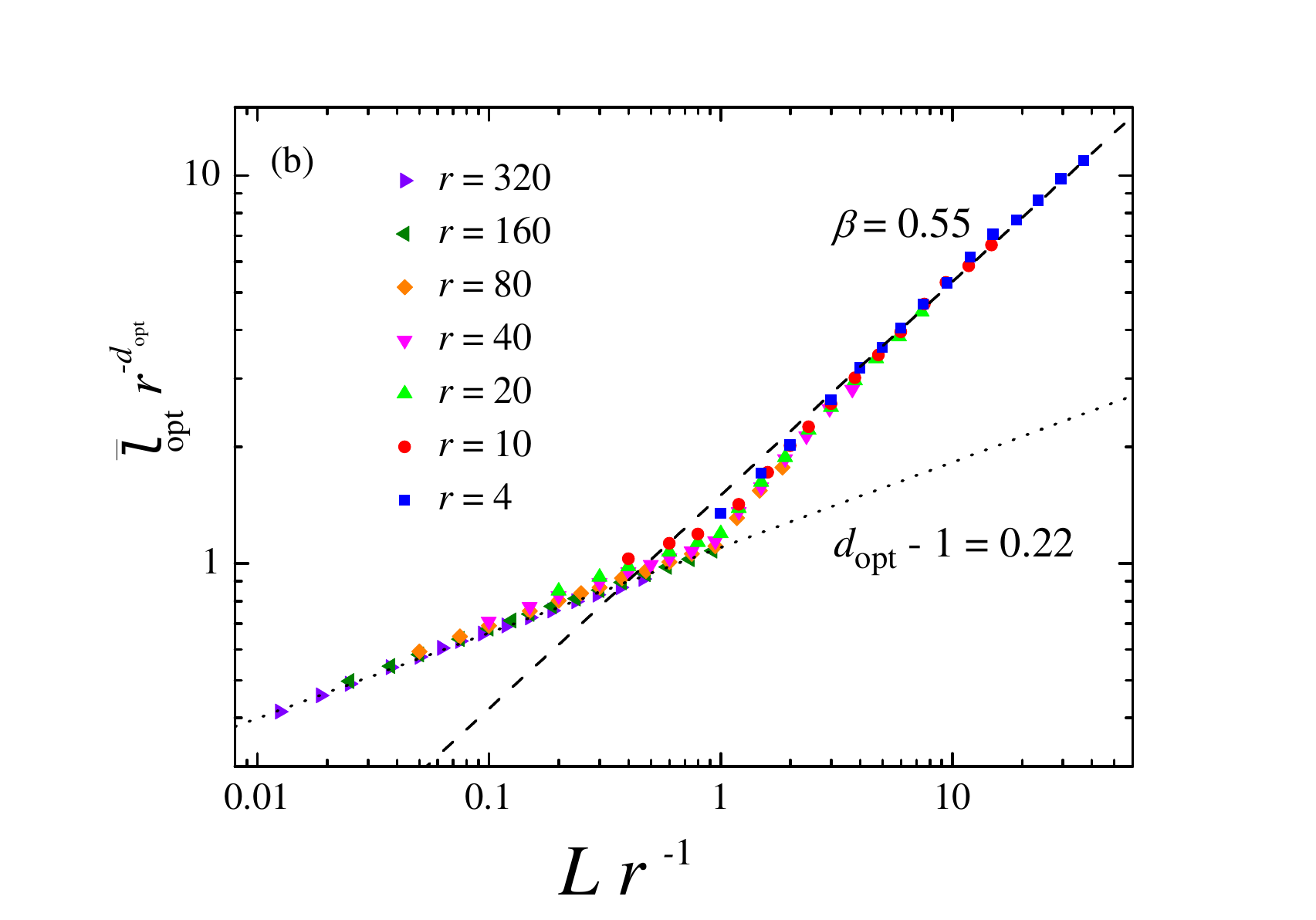}
  \caption{(a) Mean optimal path length $\loptmean$ as a function of the reduced linear size $L$ of: rectangular $r\times L$ lattices for points with $L<r$, and square $L\times L$ lattices for points with $L\geq r$, for different values of $r$ at a constant disorder strength $\Anu = 756.7$. (b) Data collapse obtained after applying the scaling given in Eq. \eqref{eq:scaling_Afijo_b}  with exponents given in Table \ref{table:scaling_exponents} for $D=2$. Broken lines in both panels represent power law behaviors with the indicated exponents.}
  \label{fig:scaling_Afijo}
\end{figure}

The next set of results show another relevant result of our model, the crossover between WD1 and  WD2, i.e., the transition that takes place in the SD limit when $L$ crosses over $r$. After applying the SD-limit conditions $r\ll \rc$ and  $L \ll \Lsat$ in Eq. \eqref{eq:general_scaling}, we obtain
\beq
\label{eq:scaling_Afijo}
\loptmean \sim  r^\alpha L^\beta \mathcal{H}\left(\frac{L}{r}\right).
\eeq
We display in Fig. \ref{fig:scaling_Afijo}(a) the scaling of $\loptmean$ with $L$ for different values of $r$. Disorder has been fixed to a constant value such that the SD-limit conditions $r \ll \Anu$ and $L \ll A^\nu$ are satisfied. For cases $L<r$ we have employed $r\times L$ rectangular lattices, whereas for $L\geq r$ we have considered square $L\times L$ lattices. Results show that, for a given value of $r$, $\loptmean$ grows as $\loptmean \sim L^{\dopt-1}$ until $L$ becomes of the same order as $r$. Then, a crossover takes place to the scaling $\loptmean \sim L^\beta$ displayed in Fig. \ref{fig:scaling_rfijo}.

We can rewrite Eq. \eqref{eq:scaling_Afijo} in the more suitable form:
\beq
\label{eq:scaling_Afijo_b}
\loptmean \sim  r^{\dopt} \mathcal{H}'\left(\frac{L}{r}\right),
\eeq
with the modified scaling function $\mathcal{H}'(x)$:
\beq
\mathcal{H}'(x)\sim \begin{cases}
                      x^{\dopt-1} & \mbox{if } x\ll 1, \\
                      x^\beta & \mbox{if } x\gg 1.
                    \end{cases}
 \eeq
In Fig. \ref{fig:scaling_Afijo}(b) we have scaled $\loptmean $ by $r^{\dopt}$, and plotted it against $L/r$, obtaining a good collapse to the scaling function $\mathcal{H}'(x)$. We observe a smooth inflexion above the crossover point $L=r$, before the scaling behavior $(L/r)^\beta$ is attained. This is probably a small effect of the lattice geometry, since we go from rectangles with points A and B located on opposite sides, to squares in which points A and B are inside the lattice.

We complete the numerical analysis of the scheme in Fig.  \ref{fig:diagrama_scalings} by addressing the transition from MD2 to WD. We thus consider $r\gg \rc$, and the argument of the scaling function $\mathcal H(x)$ is $x=L/\Anu$. From Eq. \eqref{eq:general_scaling} we have
\beq
\label{eq:scaling_M2_to_W}
\loptmean \sim  A^{\kappa-\nu}\mathcal{S}\left(\frac{L}{\Anu}\right),
\eeq
with the new scaling function
\beq
\label{eq:scaling_function_S}
 \mathcal{S}(x)=\mathcal{G}(x)\mathcal{H}(x)\sim \begin{cases}
                      x^{\dopt-1} & \mbox{if } x\ll 1, \\
                      1 & \mbox{if } x\gg 1.
                    \end{cases}
\eeq
It is again a ``Family-Vicsek''-like scaling \cite{50}, from which we obtain again the scaling relation given in Eq. \eqref{eq:scaling_relation}.

\begin{figure}
  \includegraphics[width=\columnwidth]{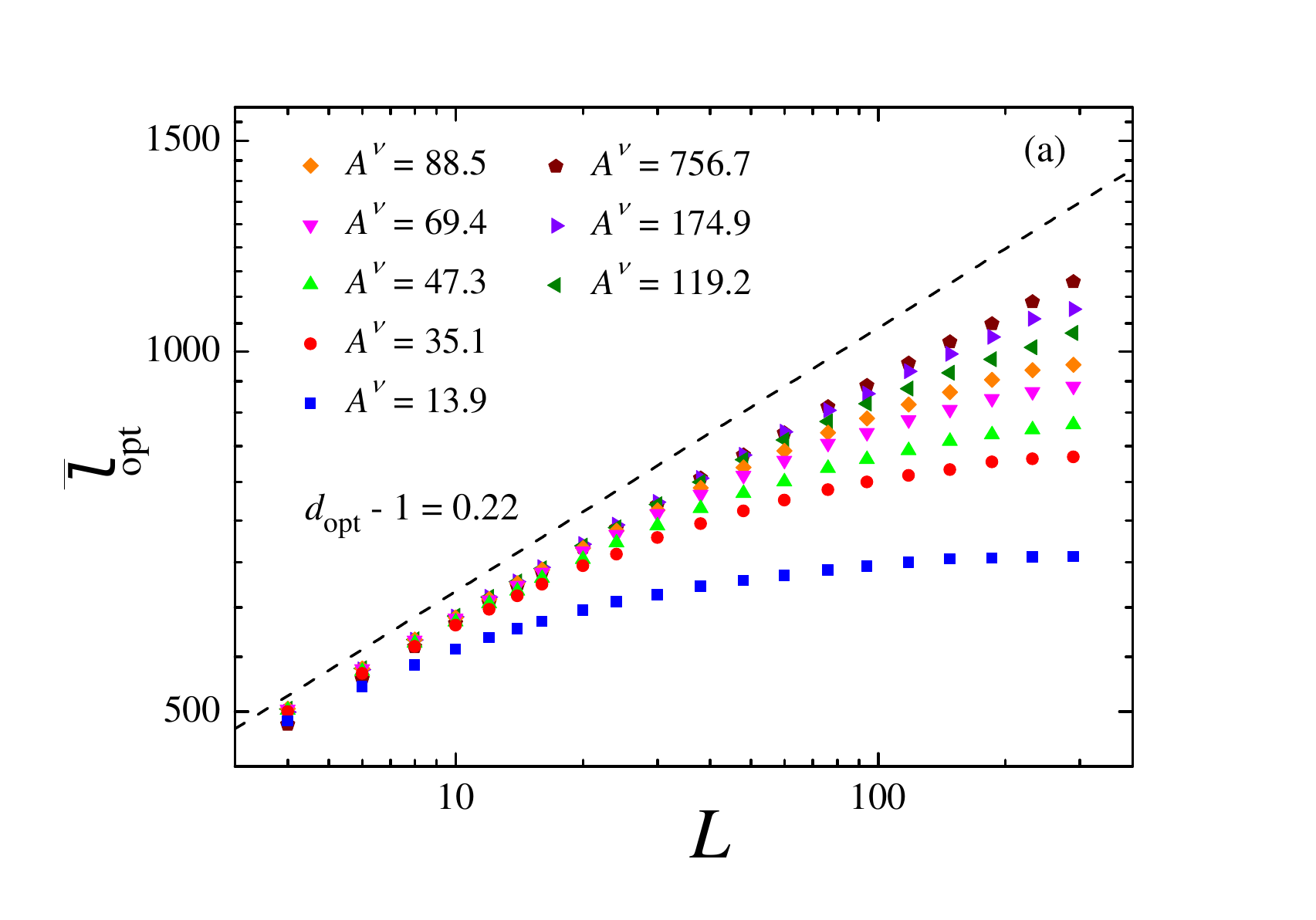}
   \includegraphics[width=\columnwidth]{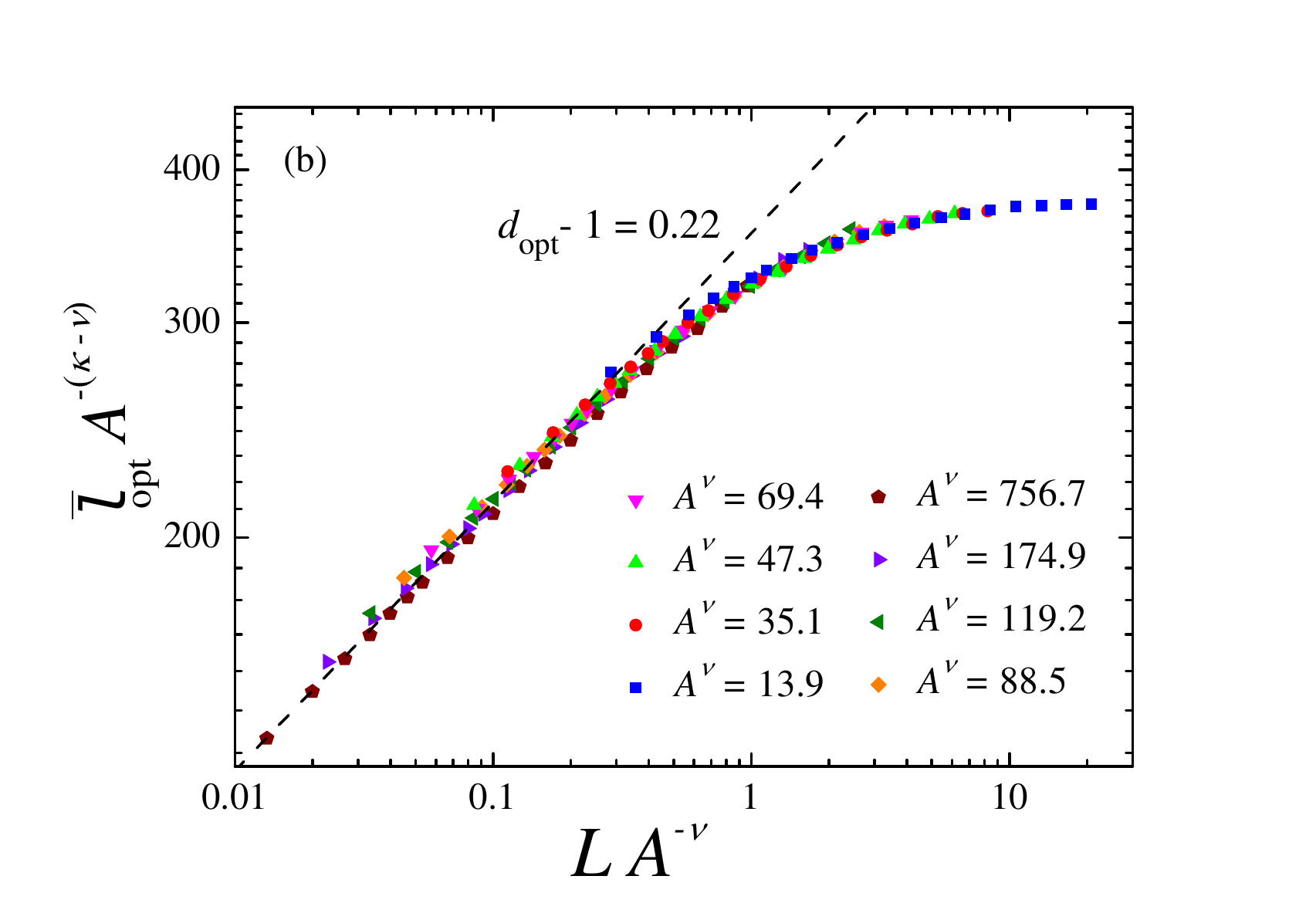}
	\caption{  (a) Mean optimal path length $\loptmean$  as a function of the reduced linear size $L$ of: rectangular $r\times L$ lattices for points with $L<r$, and square $L\times L$ lattices for points with $L\geq r$, for fixed $r=300$ and different disorder strengths. (b) Data collapse obtained after applying the scaling given in Eq. \eqref{eq:scaling_M2_to_W}  with exponents given in Table \ref{table:scaling_exponents} for $D=2$. Broken lines in both panels represent power law behaviors with the indicated exponents.}
	\label{fig:M2_to_W}
\end{figure}

We display in Fig. \ref{fig:M2_to_W}(a) the scaling of $\loptmean$ with $L$ for a fixed $r$ and different disorder strengths. Most of the disorder strengths satisfy condition $r \gg \Anu$ corresponding to case MD2, but we have also shown one case with $r \ll \Anu$ corresponding to case SD2 (both MD2 and SD2 cases have the same scaling). Points with $L<r$ are obtained using $r\times L$ rectangular lattices, and for points with $L\geq r$ we have used $L\times L$ squares. The mean optimal path length grows initially as $\loptmean \sim L^{\dopt-1}$.  When $L$ crosses over the saturation value $\Lsat$, $\loptmean$ saturates to a constant value that increases with disorder. According to the scaling given in Eq. \eqref{eq:scaling_function_S}, if we scale $L$ to the saturation value $\Anu$, and $\loptmean $ to $A^{\kappa-\nu}$, data should collapse to the scaling function $\mathcal{S}(x)$, as it is indeed displayed in Fig. \ref{fig:M2_to_W}(b).

\begin{figure}
  \includegraphics[width=\columnwidth]{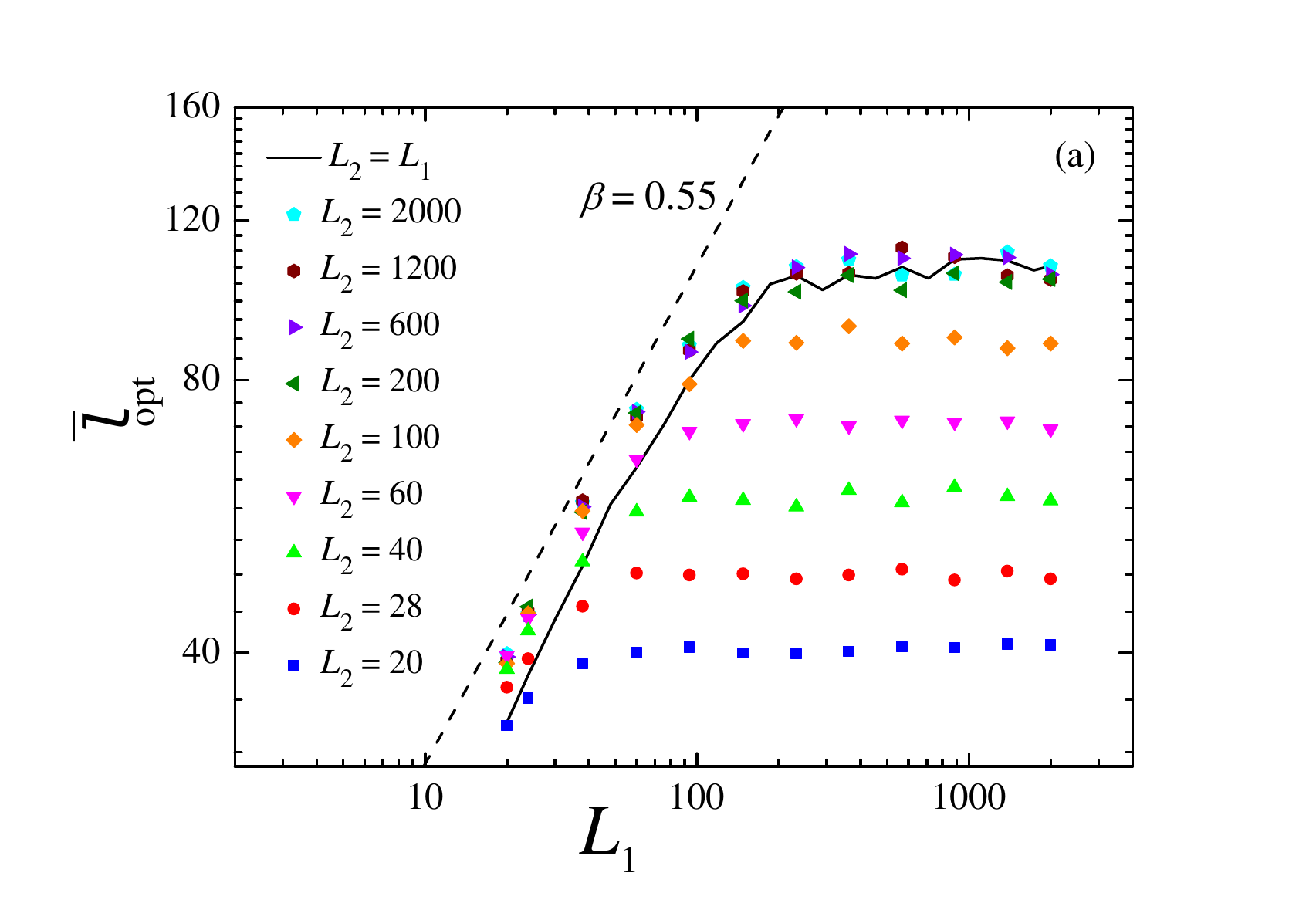}
  \includegraphics[width=\columnwidth]{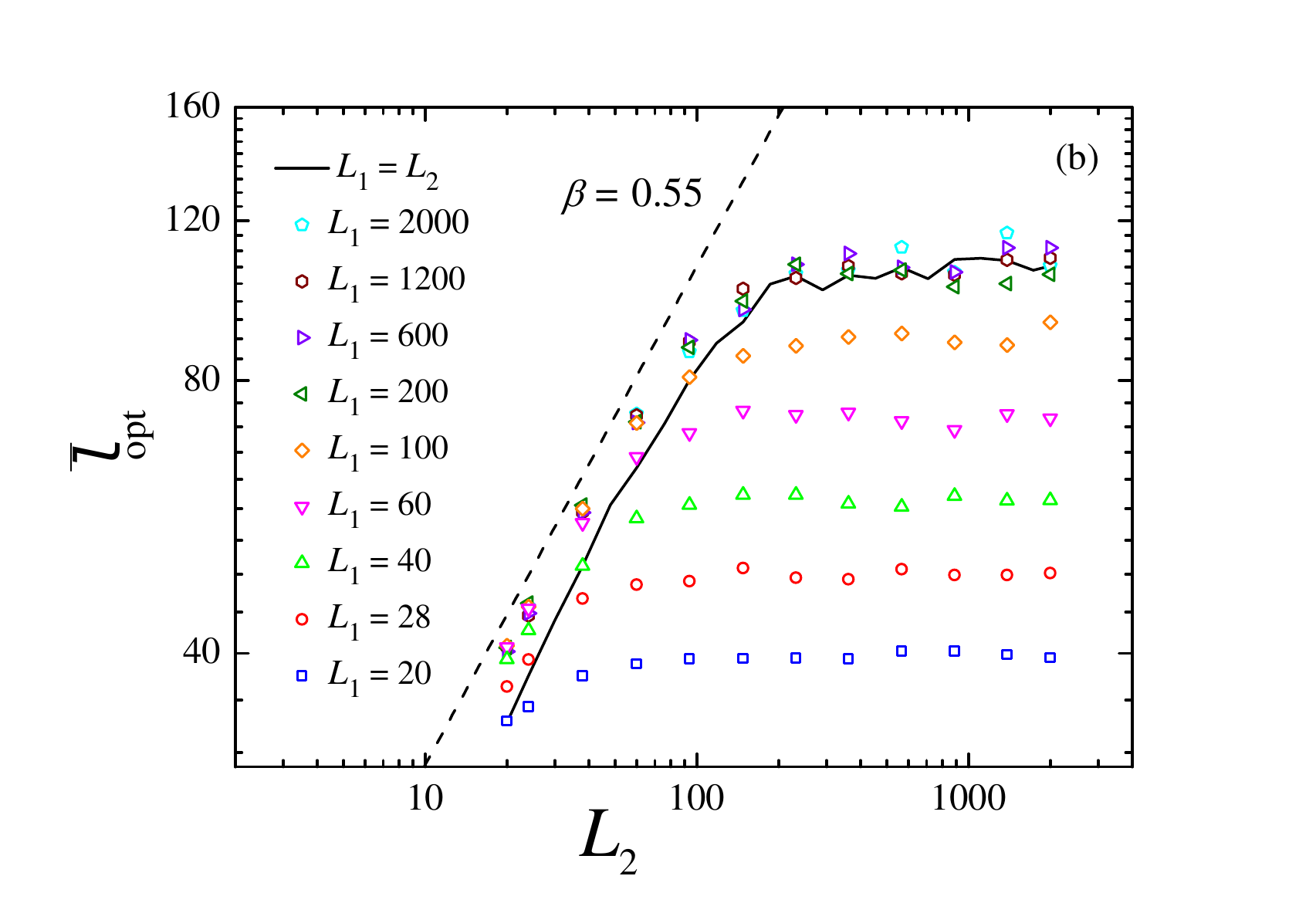}
  \caption{
  Mean optimal path length $\loptmean$ as a function of: (a) the lateral size $L_1$ of rectangular $L_1\times L_2$ lattices, for different values of $L_2$; (b) the lateral size $L_2$ of rectangular $L_1\times L_2$ lattices, for different values of $L_1$. Results correspond to $r=10$ and $\Anu=174.9$. Case $L_1=L_2=L$ displayed in Fig. \ref{fig:scaling_rfijo} has been indicated with solid thick lines. Broken lines represent power law behavior with exponent $\beta$.}
  \label{fig:scaling_Lpar_Lper}
\end{figure}

We have found implicitly in Figs. \ref{fig:scaling_Lfijo}-\ref{fig:M2_to_W} some evidences that the effects of the lattice size and its geometry are controlled by the reduced linear size $L=\min\{L_1,L_2\}$. The next set of numerical results is aimed to verify this assumption. We have repeated the case $\Anu=174.9$ displayed in Fig. \ref{fig:scaling_rfijo} but using rectangles of size $L_1\times L_2$ instead of $L\times L$ squares. We plot in Fig. \ref{fig:scaling_Lpar_Lper}(a) the scaling of $\loptmean$ with $L_1$ for different values of $L_2$, and in Fig. \ref{fig:scaling_Lpar_Lper}(b) the result of exchanging the roles of $L_1$ and $L_2$. The square lattice case displayed in Fig. \ref{fig:scaling_rfijo} has been indicated with a solid line for comparison.

The similarity between the two figures is striking, since A and B lie in both cases along the $L_1$ direction. Yet, the results indicate that both directions have the same effect. We analyze the results of Fig. \ref{fig:scaling_Lpar_Lper}(a) [the same arguing applies to Fig. \ref{fig:scaling_Lpar_Lper}(b)]. For $L_2\ll \Lsat$ ($\Lsat \approx 175$), $\loptmean$ grows with $L_1$ initially according to the expected scaling $\loptmean \sim L_1^\beta$, no matter the value of $L_2$. When $L_1$ becomes of the same order as $L_2$ (square lattice case), the mean length saturates to a value that remains constant (on average) even though $L_1$ increases above $L_2$. On the other hand, for $L_2\gg \Lsat$, the curves collapse to a single curve which coincides with the square lattice case.

\begin{figure}
  \includegraphics[width=\columnwidth]{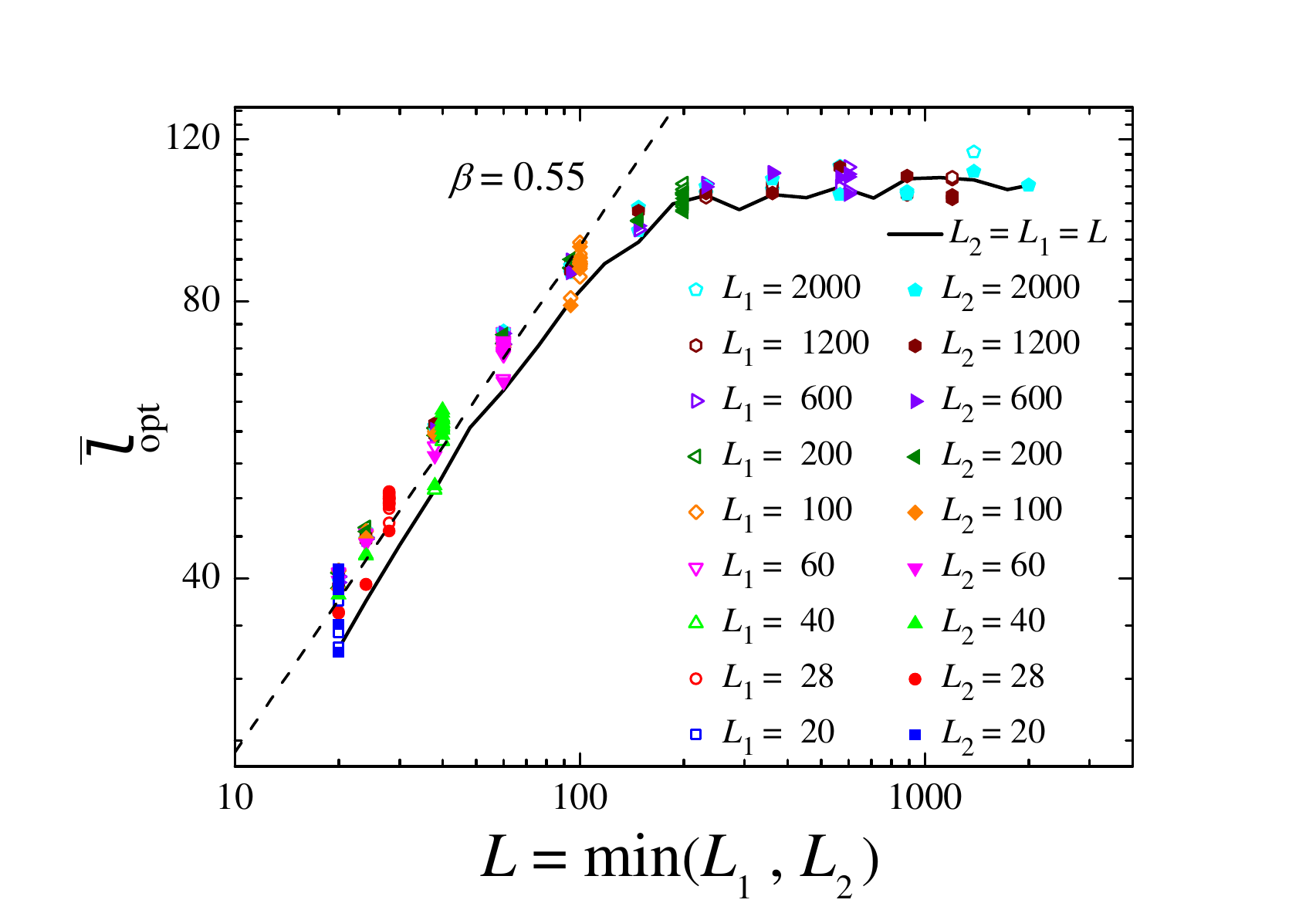}
   \caption{Same results displayed in Fig. \ref{fig:scaling_Lpar_Lper} (using the same symbols) but plotted as a function of $\min\{L_1,L_2\}$.}
  \label{fig:scaling_Lpar_Lper_colapso}
\end{figure}

There appear some deviations from this behavior when $L_2$ is small and comparable to $r$. In that case we observe a slight increase of $\loptmean$ with $L_1$ after $L_1\approx L_2$. This effect of geometry vanishes as $L_2$ increases, so for scaling purposes we can indeed assume that the relevant length scale is $L$. To support this conclusion we have plotted in Fig. \ref{fig:scaling_Lpar_Lper_colapso} all the results displayed in Fig. \ref{fig:scaling_Lpar_Lper} against the reduced linear size $L$, and we obtain an acceptable data collapse. The vertical dispersion of the points is due to the fluctuations of $\loptmean$ at saturation, so it should decrease as we increase the sampling.

\begin{figure}
  \includegraphics[width=\columnwidth]{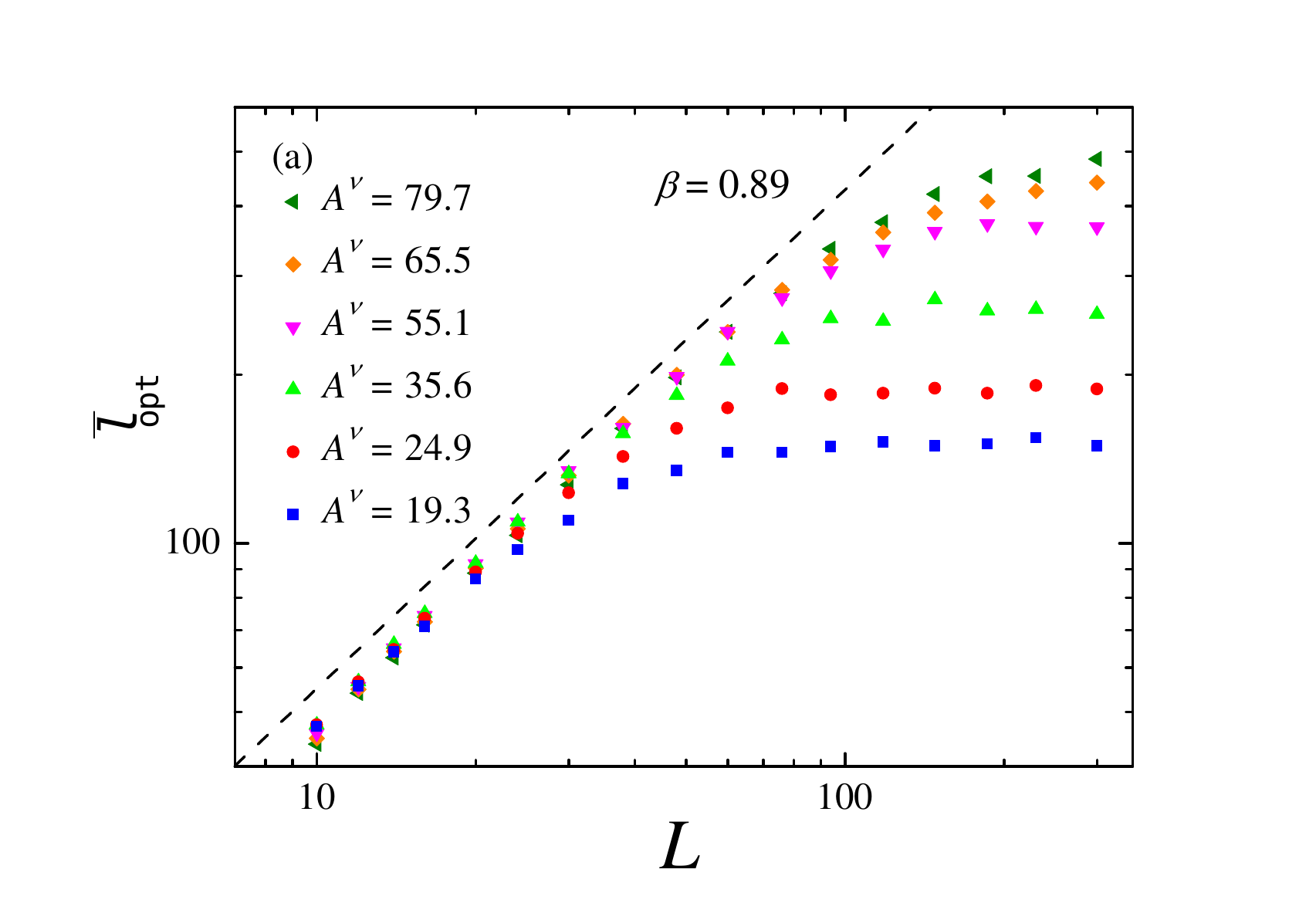}
  \includegraphics[width=\columnwidth]{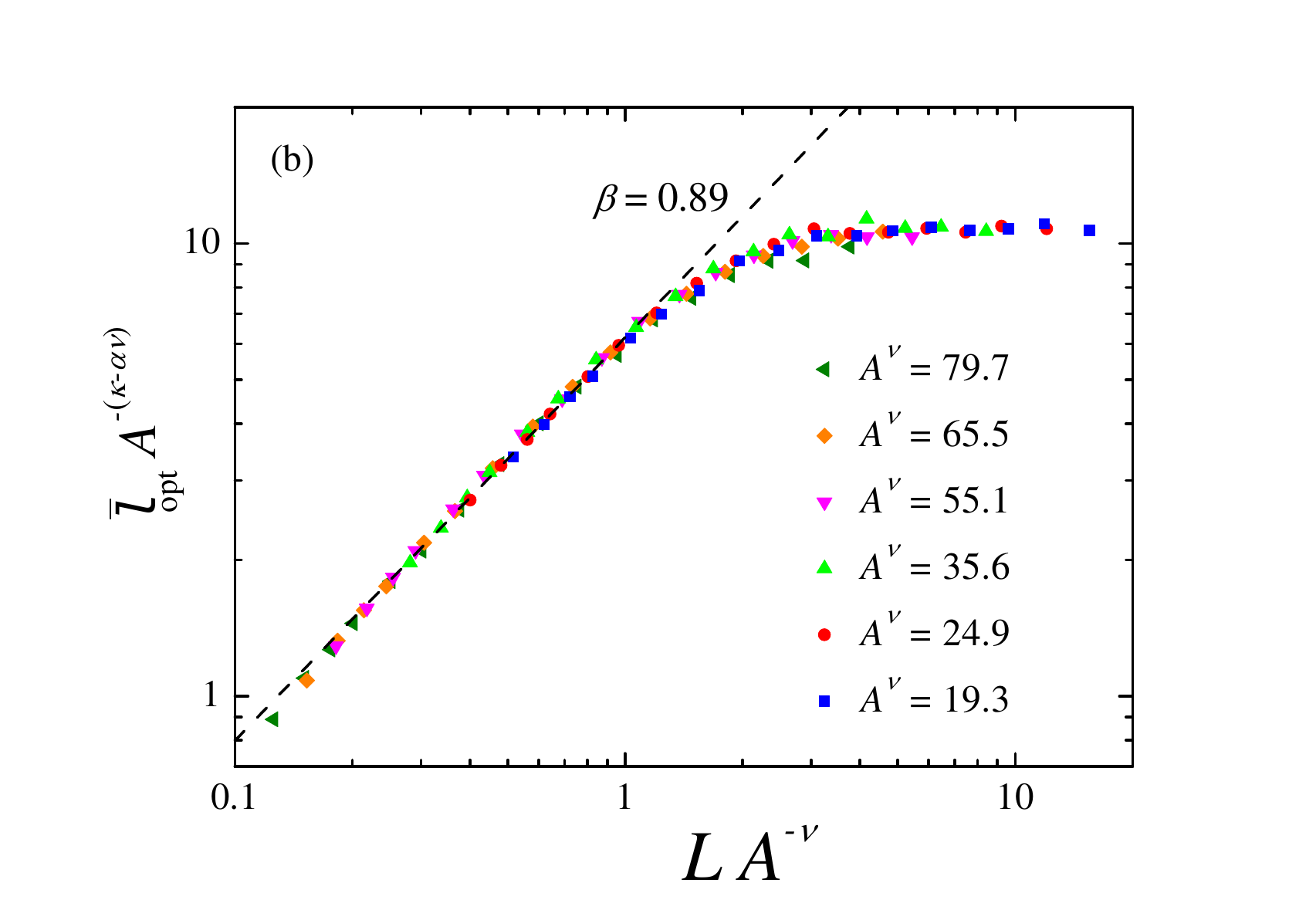}
  \caption{(a) Mean optimal path length $\loptmean$ as a function of the linear size $L$ of simple cubic lattices for $r=10$ and different disorder strengths. (b) Data collapse obtained after applying the scaling given in Eq. \eqref{eq:scaling_rfijo} with exponents given in Table \ref{table:scaling_exponents} for $D=3$. Broken lines represent power law behavior with exponent $\beta$.}

  \label{fig:scaling_rfijo_3d}
\end{figure}

\begin{figure}
  \includegraphics[width=\columnwidth]{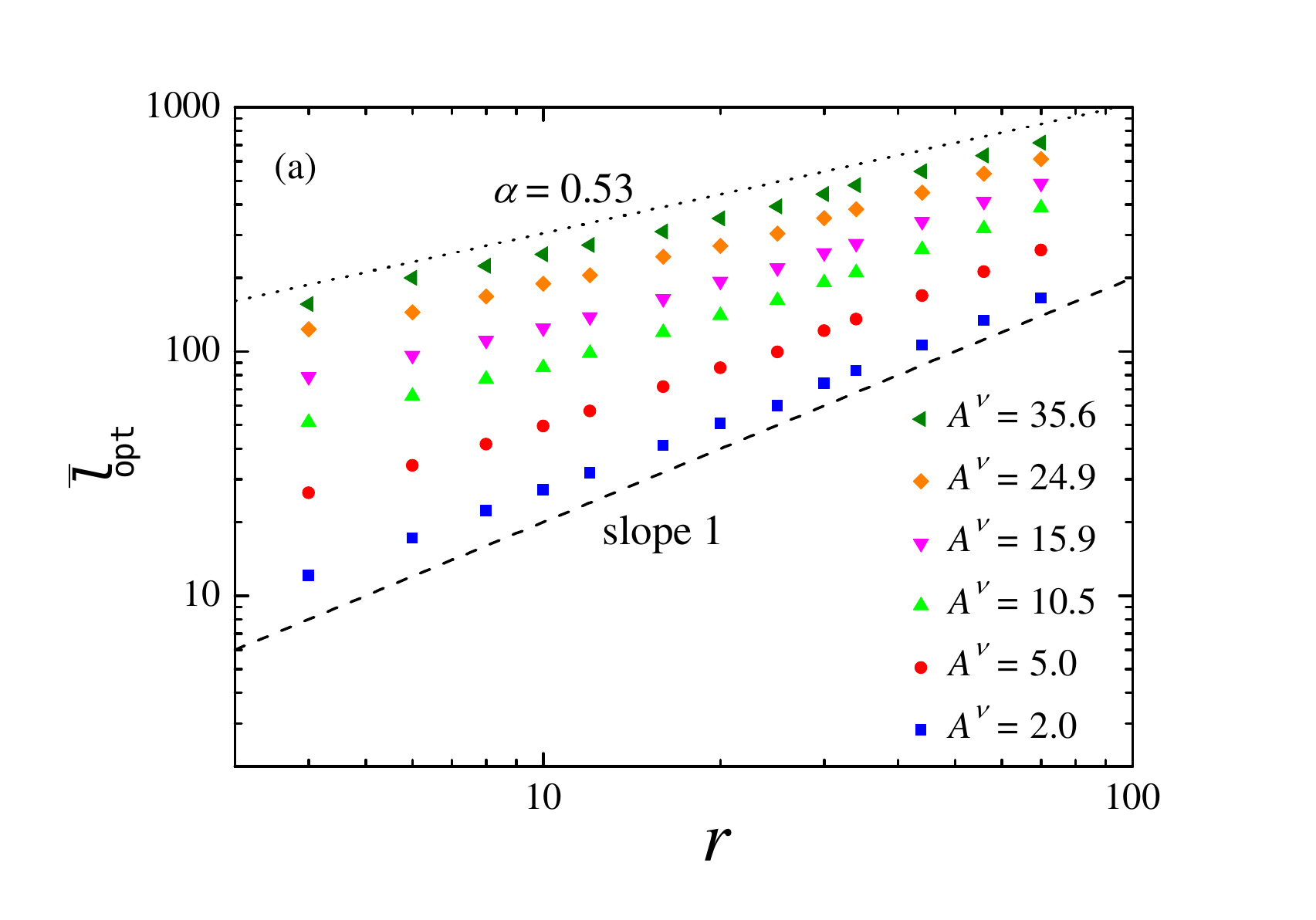}
  \includegraphics[width=\columnwidth]{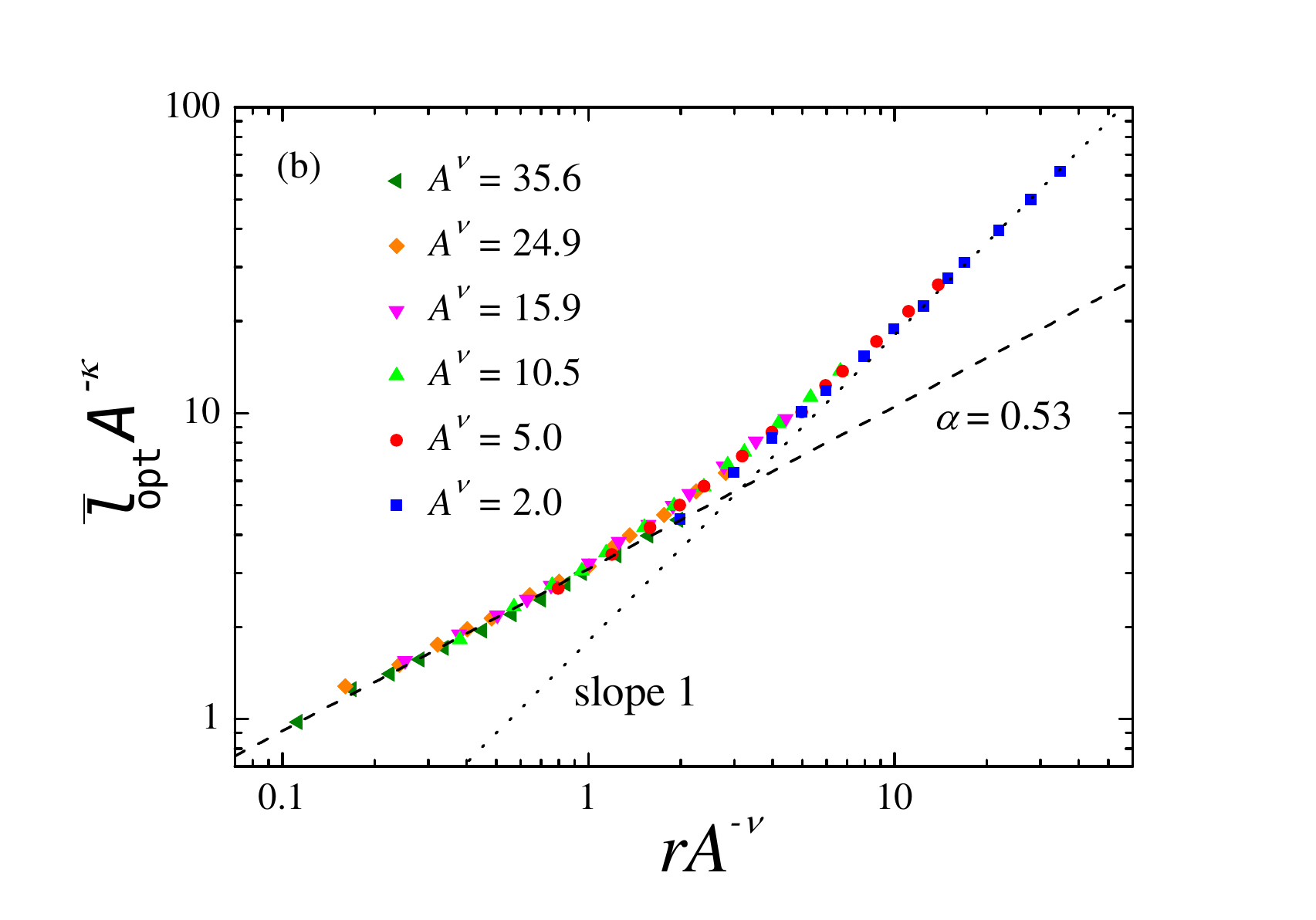}
	\caption{(a) Mean optimal path length $\loptmean$ as a function of the end-to-end distance $r$ for simple cubic lattices of constant linear size $L=100$ and different disorder strengths. (b) Data collapse obtained after applying the scaling given in Eq. \eqref{eq:scaling_Lfijo}  with exponents given in Table \ref{table:scaling_exponents} for $D=3$. In both parts broken lines represent power law behaviors with the indicated exponents.}
	\label{fig:scaling_Lfijo_3d}
\end{figure}

We finish this section by showing the results obtained in simple cubic lattices of linear size $L$. Corners are located at positions $(\pm L/2, \pm L/2, \pm L/2)$, and points A and B at $(-r/2,0,0)$ and $(r/2,0,0)$, respectively. Figure \ref{fig:scaling_rfijo_3d} is the $3D$ equivalent of Fig. \ref{fig:scaling_rfijo} and shoes the transition from SD1 to MD1 for fixed $r$. As for the $2D$ case, data displayed in Fig. \ref{fig:scaling_rfijo_3d}(a) collapse to scaling function $\mathcal{G}(x)$ when $L$ is scaled to $A^\nu$, and $\loptmean$ to $A^{\kappa -\alpha\nu}$, as displayed in Fig. \ref{fig:scaling_rfijo_3d}(b). We also show in Fig. \ref{fig:scaling_Lfijo_3d} the $3D$ equivalent of Fig. \ref{fig:scaling_Lfijo} with fixed $L=100$. Data  displayed in Fig. \ref{fig:scaling_Lfijo_3d}(a) collapse to scaling function $\mathcal{F}(x)$ when $r$ is scaled to $A^\nu$, and $\loptmean$ to $A^{\kappa}$, as displayed in Fig. \ref{fig:scaling_Lfijo_3d}(b).

\section{Conclusions and further work}
\label{sec:conclusions}

We have presented a general scaling for the mean length of the optimal path length between two points that accounts for the effects of the geometry (size and shape) of the lattice to which the points belong. It is shown that these effects are controlled by a single parameter, the smallest lateral size of the lattice, $L$, which reduces the number of degrees of freedom to three: the two length scales given by $r$ and $L$, and the disorder parameter $A$.

Interestingly, this result agrees in somehow with the \emph{invariance property} of diffusive random walks. It has been proved \cite{51}, and observed in different real systems \cite{52}, that the average length of the trajectories performed by random walks through a closed finite system, from entry point to first exit point, is given by $aV/S$, where $a$ is a numerical constant, $V$ is the system volume and $S$ its boundary. That means that it depends only on the geometry of the system. For example, for a rectangular lattice of size $L_1 \times L_2$ with $L_1 \ll L_2$, the above expression takes the form $aL_1/2$ to a good approximation, which means that the average length is controlled by the smallest linear size of the medium.

We have shown that the optimal path also experiences a strong-weak disorder transition with respect to $L$, when $L$ reaches a certain characteristic length $\Lsat$ that scales with disorder in the same way as the crossover end-to-end distance, $\Lsat \sim \rc\sim \Anu$. The numerical results presented here indicate that the numerical prefactors in the scaling relations for $\rc$ and $\Lsat$ are nearly the same. They are approximately equal to 1 in $D=2$, and above 2 in $D=3$. However, there is remarkable difference between these two disorder-induced transitions. In the case of $L$, the mean length saturates to a constant value, whereas for $r$, it experiences a crossover.

The general scaling proposed here considers three scaling exponents (besides the percolation connectivity exponent $\nu$). The exponent $\alpha$ determines the power-law growth of $\loptmean$ with $r$ in a fixed lattice with SD-limit conditions with respect to $r$. The exponent $\beta$ determines the power-law growth of $\loptmean$ with $L$, for a fixed $r$, in the SD limit. The exponent $\kappa$ characterizes the scaling, with the disorder strength, of the mean optimal path length at both saturation and crossover points. The three exponents are related through the ``Family-Vicsek''-like scaling relation $\kappa =\nu(\alpha+\beta)$.

The decoupling of $r$ and $L$ allowed us to go deeper into the origin of the universal scaling in the SD limit. In particular, we deduced two scaling relations between our scaling exponents and universal exponents $\dopt$ and $\gopt$: $\dopt=\alpha+\beta$ and $\gopt=1+\alpha/\dopt$. We showed how the scaling behaviors reported in the literature arise naturally from our model as particular cases, and we completed the casuistry of the problem by addressing the new scaling regimes revealed by the model.

The next steps in future work should focus on elucidating the dependence of the scaling exponents on the dimension $D$ of the system. It has been conjectured \cite{20,36} that $\dopt$ monotonically increases with $D$, from $\dopt=1$ in $D=1$, to $\dopt=2$ for $D\geq D_c=6$, which is the upper critical dimension of percolation \cite{46}. For $D\geq D_c$ the optimal path has a fractal dimension of 2 corresponding to a random walk. On the other hand, we have that $\gopt < 2$ for $D>1$, and it seems to decrease with $D$ \cite{35}. In addition, for $L=\infty$, the integrability condition imposes condition $\gopt > 1$ .

We can now assume that the scaling relation $\alpha=\dopt/D$ derived from our analysis of the radius of gyration, and whose verification certainly deserves additional work, holds for $D\leq d_c$. After considering all the above conditions, we can conjecture the following: (i) $\alpha$ decreases with $D$ according to $\alpha=\dopt/D$, ranging from $\alpha=1$ in $D=1$ to $\alpha=1/3$ in $D=D_c$; (ii) $\beta$ increases with $D$ following the relation $\beta=\dopt(1-D^{-1})$, from $\beta=0$ in $D=1$ up to $\beta=5/3$ in $D=D_c$; (iii) exponent $\gopt$ decreases with $D$ according to $\gopt=1+D^{-1}$, taking the value $\gopt=7/6$ in  $D=D_c$.  As discussed above, more research is needed in this regard. Another issue of great importance that also deserves additional work is obtaining accurate values of the exponents $\alpha$ and $\beta$.

\section*{Acknowledgments}
\label{sec:Acknowledgments}

This work was partially supported by Grants PGC2018-094763-B-I00 and PID2021-123969NB-I00 funded by MCIN/AEI/10.13039/501100011033 and by ``ERDF A way of making Europe''. D.V. acknowledges Grant PRE2019-088226 funded by MCIN/AEI/10.13039/501100011033 and by ``ERDF A way of making Europe''.
We acknowledge the computational resources and assistance provided by the Centro de Computaci\'on de Alto Rendimiento CCAR-UNED.  Authors also thank  J. Rodr\'{\i}guez-Laguna and S.N. Santalla for fruitful discussions.


\end{document}